\documentclass[a4paper,11pt]{article}
\pdfoutput=1 

\usepackage{jheppub} 
\usepackage{etoolbox}
\makeatletter
    \patchcmd{\maketitle}{\@fpheader}{}{}{}
    \makeatother
    
\usepackage[T1]{fontenc} 
\usepackage{subcaption}
\usepackage{comment}

\title{Aspects of CFTs on Real Projective Space}

\author[]{Simone Giombi$^{a}$,}
\author[]{Himanshu Khanchandani$^{a}$,}
\author[]{Xinan Zhou$^{b}$}
\affiliation[]{$^{a}$Department of Physics, Princeton University, Princeton, NJ 08544, USA }
\affiliation[]{$^{b}$Princeton Center for Theoretical Science, Princeton University, Princeton, NJ 08544, USA }

\abstract{We present an analytic study of conformal field theories on the real projective space $\mathbb{RP}^d$, focusing on the two-point functions of scalar operators. Due to the partially broken conformal symmetry, these are non-trivial functions of a conformal cross ratio and are constrained to obey a crossing equation. After reviewing basic facts about the structure of correlators on $\mathbb{RP}^d$, we study a simple holographic setup which captures the essential features of boundary correlators on $\mathbb{RP}^d$. The analysis is based on calculations of Witten diagrams on the quotient space $AdS_{d+1}/\mathbb{Z}_2$, and leads to an analytic approach to two-point functions. In particular, we argue that the structure of the conformal block decomposition of the exchange Witten diagrams suggests a natural basis of analytic functionals, whose action on the conformal blocks turns the crossing equation into certain sum rules. We test this approach in the canonical example of $\phi^4$ theory in dimension $d=4-\epsilon$, extracting the CFT data to order $\epsilon^2$. We also check our results by standard field theory methods, both in the large $N$ and $\epsilon$ expansions. Finally, we briefly discuss the relation of our analysis to the problem of construction of local bulk operators in AdS/CFT.}

\begin{document}
\maketitle
\flushbottom
\section{Introduction}
Quantum field theories on non-orientable spacetime have several physical applications, and have been studied from many different perspectives. They are an integral part of string theory in the description of unoriented worldsheets \cite{Polchinski:1998rq,Blumenhagen:2009zz,Fioravanti:1993hf}.  Studying theories on non-orientable manifolds also probes the realization of time reversal symmetry \cite{Kapustin:2014tfa,Kapustin:2014gma,Kapustin:2014dxa} (see, {\it e.g.}, \cite{Metlitski:2015yqa} for discussions on a refinement of electric-magnetic duality in abelian gauge theories, and \cite{Guo:2017xex,Wan:2018zql,Wan:2019oyr,Wang:2019obe} on Yang-Mills theory and $\mathbb{CP}^{N-1}$-sigma models), which plays important roles in condensed matter physics. They also  make appearances in formal studies of quantum field theory. For example, partition functions of two dimensional CFTs on non-orientable surfaces were studied in \cite{Maloney:2016gsg}, where holographic connections with three dimensional geometries were explored. Their role in supersymmetric quantum field theories was discussed, {\it e.g.}, in \cite{LeFloch:2017lbt, Wang:2020jgh}. Recently, there has been considerable interest in studying CFTs on real projective space -- one of the simplest examples of non-orientable manifolds.\footnote{More precisely, $\mathbb{RP}^d$ is unorientable for $d$ even, and orientable for $d$ odd.}  This is partly in light of the modern nonperturbative conformal bootstrap  (see \cite{Rychkov:2016iqz,Poland:2018epd} for reviews)  \cite{Nakayama:2016cim, Hasegawa:2016piv,Hogervorst:2017kbj, Hasegawa:2018yqg}, where CFTs on real projective space provide  attractive playgrounds for developing and testing new techniques. 
Moreover, the study of such theories is also fueled by the program of constructing bulk local operators in AdS \cite{Miyaji:2015fia,Nakayama:2015mva, Verlinde:2015qfa, Nakayama:2016xvw,Goto:2016wme, Lewkowycz:2016ukf}, where crosscap states are proposed to be dual to fields inserted at a bulk point. 
In this paper, we continue the analytic study of conformal field theories on real projective space, and revisit the problem from multiple angles.

The real projective space $\mathbb{RP}^d$ can be defined by a $\mathbb{Z}_2$ quotient of a sphere $S^d$
\begin{equation}
\mathbf{X}^2=1\;,\quad \mathbf{X}\in \mathbb{R}^{d+1}\;,\quad \text{with} \;\; \mathbf{X}\sim -\mathbf{X}.
\end{equation}
Equivalently, since our focus is on CFT, we can perform a Weyl transformation and map it to the flat space
\begin{equation} \label{MetricWeyl}
x^{\mu} = \frac{X^{\mu}}{1 - X^{d + 1}}, \ \mu=1,\ldots,d \hspace{1cm} ds^2_{\mathbb{R}^d} = \frac{(1 + x^2)^2}{4} ds^2_{S^d}\;.
\end{equation} 
The real projective space is then represented as $\mathbb{R}^d$ under the identification 
\begin{equation}
\label{inversion}
x^\mu\to -\frac{x^\mu}{x^2}
\end{equation}
where $x^\mu$ are the Cartesian coordinates on $\mathbb{R}^d$. Unless otherwise stated, in this paper we denote by $\mathbb{RP}^d$ the quotient of flat space by the inversion (\ref{inversion}).  

Putting CFTs on $\mathbb{RP}^d$ partially breaks the conformal symmetry and introduces new observables. Scalar operators can have non-vanishing one-point functions 
\begin{equation}
\langle \mathcal{O}_\Delta \rangle=\frac{a_{\mathcal{O}}}{(1+x^2)^{\Delta_{\mathcal{O}}}}\;.
\end{equation}
The coefficients $a_{\mathcal{O}}$ are new data that defines the CFT on real projective space, along with the standard operator spectrum and OPE coefficients, which remain the same as on $\mathbb{R}^d$. Moreover, two-point functions are no longer fixed by symmetry, but instead become functions of a cross ratio $\eta$ invariant under the residual conformal symmetry
\begin{equation}
\langle \mathcal{O}_1(x_1)\mathcal{O}_2(x_2)\rangle=\frac{\mathcal{G}(\eta)}{(1+x_1^2)^{\Delta_1}(1+x_2^2)^{\Delta_2}}\;,
\end{equation}
where
\begin{equation}
\eta=\frac{(x_1-x_2)^2}{(1+x_1^2)(1+x_2^2)}\;.
\end{equation}
The reader might notice the breaking of conformal symmetry and the structure of correlators are very reminiscent of those of boundary CFTs \cite{McAvity:1995zd,Liendo:2012hy}. We will point out more similarities later in the paper. Analogous  to four-point functions on $\mathbb{R}^d$, two-point functions of $\mathbb{RP}^d$ CFTs obey a crossing equation because of the identification (\ref{inversion})
\begin{equation}\label{crossingintro}
\mathcal{G}(\eta)=\pm\, \mathcal{G}(1-\eta)
\end{equation}
where $\pm$ corresponds to the two choices $\mathcal{O}_{1,2} \to \pm\, \mathcal{O}_{1,2}$ under the inversion. The operator product expansion in the direct channel ($\eta\to 0$), and in the mirror channel ($\eta\to 1$) allows them to be expanded in terms of conformal blocks in the respective channels. The crossing equation together with the conformal block decomposition then impose nontrivial constraints on the structure of correlators. Two-point functions are therefore the prime target for developing an analytic understanding of CFTs on real projective space. 

In this paper, we develop an analytic approach to two-point functions, which is universal for $\mathbb{RP}^d$ CFTs. However, to develop this method we will take a holographic detour. We first lift  the quotient (\ref{inversion}) into the bulk as a $\mathbb{Z}_2$ quotient of $AdS_{d+1}$, and study a toy model for holography in this setup. 
We define the tree-level Witten diagrams in this background, and study their various properties. In particular, we  consider in detail the two-point conformal block decompositions of exchange Witten diagrams in the two channels. The structure of the conformal block decompositions suggests a natural basis for the function space of two-point correlators, which consists of special conformal blocks with discrete `double-trace' dimensions in both the direct and the mirror channel. The dual of this basis is a basis of analytic functionals, whose actions on a generic conformal block can be read off from the conformal block decomposition coefficients of the exchange Witten diagrams. Acting on the crossing equation (\ref{crossingintro}) with the functionals allows us to extract the complete set of constraints in the form of sum rules. These sum rules are valid non-perturbatively. But they become especially simple around the mean field theory spectrum, and essentially trivialize the study of perturbations around mean field theory. We demonstrate the use of the analytic functionals on the model of $\phi^4$ theory in $4-\epsilon$ dimensions. By solving the functional sum rules, we obtain the one-point function coefficients to the order $\epsilon^2$. Setting $\epsilon=1$, we find good agreement with the numerical bootstrap estimation for the 3d Ising model \cite{Nakayama:2016cim}.

We also develop perturbative field theory approaches to study CFTs on $\mathbb{RP}^d$, with the $O(N)$ vector model being our main example. We study the two-point function of the fundamental scalar $\phi$ of the $O(N)$ model both in the large $N$ expansion in arbitrary dimension, and in the $\epsilon$ expansion in $d = 4 - \epsilon$ dimensions. In $d = 4- \epsilon$, instead of using the usual loop expansions, we exploit the fact that $\phi$ satisfies an equation of motion. The equation of motion implies a differential equation obeyed by the two-point function, which can be solved in perturbation theory to order $\epsilon^2$. The essential idea of using equations of motion to obtain CFT data was described in \cite{Rychkov:2015naa} and here we extend it to CFTs on $\mathbb{RP}^d$. The field theory calculations provide an independent test of the results obtained from analytic functionals. 

We also discuss other interesting features of $\mathbb{RP}^d$ CFT. We point out a two-term `dimensional reduction' formula for conformal blocks, which expresses a conformal block in $d-2$ dimensions as the sum of two $d$ dimensional conformal blocks with shifted conformal dimensions. An analogous five-term relation was found in \cite{Kaviraj:2019tbg} for CFT four-point functions in $\mathbb{R}^d$, and was shown to be a consequence of the Parisi-Sourlas supersymmetry \cite{Parisi:1979ka}. The appearance of the dimensional reduction relation therefore suggests a possible extension of the Parisi-Sourlas supersymmetry to real projective space. Moreover, following \cite{Zhou:2020ptb}, we show that the dimensional reduction formula for conformal blocks can be extended to exchange Witten diagrams as well. 

The setup of our toy model for holography is also closely related to the Hamilton-Kabat-Lifschytz-Lowe (HKLL) approach for constructing local bulk operators \cite{Hamilton:2005ju,Hamilton:2006fh,Hamilton:2006az}. The two problems have the same partially broken conformal symmetry. We will make a few comments on how our results are relevant in the bulk reconstruction problem. In particular, we point out that the bulk reconstruction of the bulk-boundary-boundary three-point function can be reformulated as a conformal bootstrap problem, which can be solved using our functionals. 

The rest of the paper is organized as follows. In Section \ref{Sec:2}, we review the kinematics of $\mathbb{RP}^d$ CFT  using the embedding formalism. We set up the holography toy model on the $\mathbb{Z}_2$ quotient of AdS in Section \ref{Sec:3}, and define various Witten diagrams. In Section \ref{Sec:4} we perform a detailed analysis of the tree-level two-point exchange Witten diagrams: we evaluate them in a closed form and study their conformal block decompositions. In Section \ref{Sec:5} we study the dimensional reduction of conformal blocks and exchange Witten diagrams. We develop a functional method to $\mathbb{RP}^d$ CFTs in Section \ref{Sec:6}, and use the method in a few perturbative applications. We also present a complementary field theory approach using the equation of motion method. The relation to bulk reconstruction is discussed in Section \ref{Sec:7}. We conclude in Section \ref{Sec:8} with a brief discussion of future directions. In Appendix \ref{AppendixFreeEnergy}, we compute the free energy on $\mathbb{RP}^d=S^d/\mathbb{Z}_2$ for $O(N)$ models; the calculation has some connection to the content of Section \ref{Sec:CFTEoM}, but is mostly independent from the main text of the paper and can be read separately.

\section{Kinematics of CFT on $\mathbb{RP}^d$}\label{Sec:2}
\subsection{Embedding space}\label{Sec:2.1}
It is convenient to introduce the embedding space which linearizes the action of the conformal group. Let us first review the case where the space is just $\mathbb{R}^{d-1,1}$. For any point $x^\mu\in\mathbb{R}^{d-1,1}$, we can represent it as a null ray $P^A$ ($A=1,2,\ldots,d+2$) in $\mathbb{R}^{d,2}$
\begin{equation}
P\cdot P=0\;,\quad P\sim \lambda\, P\;.
\end{equation}
Operators are defined on the space of null rays\footnote{For simplicity we focus here on scalar operators.}, with the scaling property
\begin{equation}\label{scaling}
\mathcal{O}_\Delta(\lambda P)=\lambda^{-\Delta} \mathcal{O}_\Delta(P)\;.
\end{equation}
For definiteness, let us choose the signature of $\mathbb{R}^{d,2}$ to be $(-,+,-,+,\ldots,+)$.\footnote{For Euclidean spacetime $\mathbb{R}^d$ the embedding space is $\mathbb{R}^{d+1,1}$, and we choose the signature to be $(-,+,+,+,\ldots,+)$.} We can explicitly parameterize $P^A$ with the $\mathbb{R}^{d-1,1}$ coordinates as
\begin{equation}
P^A=\left(\frac{1+x^2}{2},\frac{1-x^2}{2},x^\mu\right)\;,
\end{equation}
 after fixing a gauge for the rescaling freedom
\begin{equation}\label{gaugefixing}
P^1=\frac{1}{2}(1+x^2)\;.
\end{equation}
 The conformal group $SO(d,2)$ acts on $P^A$ as linear rotations in $\mathbb{R}^{d,2}$
 \begin{equation}
 P^A\to \Omega^A_B\, P^B\;.
 \end{equation}
The conformal transformation on the $x^\mu$ coordinates is obtained after restoring the gauge fixing condition (\ref{gaugefixing}) by an appropriate rescaling. 

The inversion (\ref{inversion}) can be conveniently represented in terms the embedding coordinates, where it flips the sign of the last $d+1$ components of the embedding vector
\begin{equation}\label{calIembed}
\mathcal{I}: P^1\to P^1\;,\quad P^a\to -P^a\;,\quad a=2,3,\ldots,d+2\;.
\end{equation}
To go back to (\ref{gaugefixing}) we must multiply the null vector with a factor $x^{-2}$, and one can easily check that this reproduces the transformation (\ref{inversion}). Under inversion operators are identified by 
\begin{equation}\label{opidentification}
\mathcal{I}: \mathcal{O}^\pm_\Delta(x)\to \pm\, x^{2\Delta}\mathcal{O}^\pm_\Delta (x')\;,\quad x'^\mu=-\frac{x^\mu}{x^2}\;.
\end{equation}

The insertion of a crosscap introduces a fixed {\it time-like} embedding vector 
\begin{equation}\label{Nc}
N_c=(1,0,0,\ldots,0)\;.
\end{equation}
The residual conformal symmetry after inserting the crosscap is all the $SO(d,1)$ rotations that leaves $N_c$ invariant. It is useful to compare the situation with the closely related case of CFTs with a conformal boundary. The presence of a spherical boundary centered at $x=0$ with unit radius, is represented in the embedding space by introducing a {\it space-like} constant vector 
\begin{equation}\label{NB}
N_B=(0,1,0,0,\ldots,0)\;.
\end{equation}
The conformal boundary breaks the $SO(d,2)$ conformal group down to the subgroup $SO(d-1,2)$, which consists of all the rotations in the embedding space preserving the vector $N_B$.

\subsection{Correlators and conformal blocks}
The embedding space formalism introduced in the last section makes it straightforward to discuss the kinematics of CFT correlators on $\mathbb{RP}^d$. Correlators are constructed using the $SO(d,1)$ invariants of the embedding vectors, and must scale properly according to (\ref{scaling}).

Let us start with one-point functions. The only invariant one can write down is $(-2 N_c\cdot P)$, and scaling requires the one-point function must be of the form \footnote{Throughout this paper, we normalize the correlation functions by the $\mathbb{RP}^d$ partition function so that the one-point function of the identity operator is 1. In addition, we also normalize the operators such that in the short distance limit, the two-point function goes as $ \langle \mathcal{O}_{\Delta} (x_1)\mathcal{O}_{\Delta} (x_2) \rangle \sim \frac{1}{(x_1 - x_2)^{2 \Delta}}$. } 
\begin{equation}\label{1ptfun}
\langle \mathcal{O}_\Delta(x)\rangle=\frac{a_{\mathcal{O}}}{(-2N_c\cdot P)^\Delta}=\frac{a_{\mathcal{O}}}{(1+x^2)^\Delta}\;.
\end{equation}
Using the Weyl transformation (\ref{MetricWeyl}), this implies that on the $\mathbb{Z}_2$ quotient of the sphere, the one-point functions are constant
\begin{equation}\label{1ptfunSphere}
\langle \mathcal{O}_{\Delta}(x) \rangle_{\mathbb{S}^d/ \mathbb{Z}_2} = \frac{a_{\mathcal{O}}}{2^{\Delta}}.
\end{equation} 
 Note that under the inversion (\ref{opidentification}), the operator $\mathcal{O}$ must transform with the $+$ sign in order for the one-point function to be nonvanishing. This can be clearly seen from the fact that the one-point function is a constant on the sphere ${\mathbb{S}^d/ \mathbb{Z}_2}$, and we note that antipodal points on the sphere are identified by inversion. The choice of the $-$ sign leads to a zero value for $a_{\mathcal{O}}$. More generally, it follows from the fact that the total $\mathbb{Z}_2$ charge under inversion must be zero in a correlator. Therefore, one-point functions are completely determined by symmetry up to a constant $a_{\mathcal{O}}$. The constant $a_{\mathcal{O}}$ is a new CFT data, and encodes dynamical information of CFTs on real projective space. We should also point out that only scalar operators can obtain nonzero one-point function. Operators with spin must have vanishing one-point functions, because a nonzero one-point function is inconsistent with the residual symmetry (a completely analogous result holds in BCFT).   

We now discuss two-point functions, which are the main focus of this paper. With the embedding vectors $P_1$, $P_2$ and $N_c$, we can construct a cross ratio
\begin{equation}\label{defeta}
\eta=\frac{(-2 P_1\cdot P_2)}{(-2N_c\cdot P_1)(-2N_c\cdot P_2)}=\frac{(x_1-x_2)^2}{(1+x_1^2)(1+x_2^2)}\;,
\end{equation}
which is also invariant under the independent rescaling of each operator. In Euclidean spacetime, the cross ratio takes values in $\eta\in[0,1]$. After extracting a kinematic factor which takes care of the scaling property, we can write the two-point function as a function of the cross ratio
\begin{equation}
\langle \mathcal{O}_{\Delta_1}(x_1)\mathcal{O}_{\Delta_2}(x_2)\rangle=\frac{\mathcal{G}(\eta)}{(-2N_c\cdot P_1)^{\Delta_1}(-2N_c\cdot P_2)^{\Delta_2}}=\frac{\mathcal{G}(\eta)}{(1+x_1^2)^{\Delta_1}(1+x_2^2)^{\Delta_2}}\;.
\end{equation}
Here we have suppressed the parity of the operators under $\mathbb{Z}_2$. The correlator is only nonzero when $\mathcal{O}_{\Delta_1}$ and $\mathcal{O}_{\Delta_2}$ have the {\it same} parity, in order for the correlator to have a zero overall $\mathbb{Z}_2$ charge.
Moreover, because of the operator identification (\ref{opidentification}) the correlator $\mathcal{G}(\eta)$ must satisfy the following {\it crossing equation} \cite{Nakayama:2016cim}
\begin{equation}\label{crossingeqn}
\mathcal{G}(\eta)=\pm\, \mathcal{G}(1-\eta)\;,
\end{equation}
where $\pm$ denotes the common parity of $\mathcal{O}_{\Delta_1}$ and $\mathcal{O}_{\Delta_2}$. Here and elsewhere, the upper sign refers to the $+$ parity and the lower sign to $-$ parity. Also note again that under the Weyl transform \eqref{MetricWeyl}, the two-point function on ${\mathbb{S}^d/ \mathbb{Z}_2}$  just becomes $\mathcal{G} (\eta)/2^{\Delta_1 + \Delta_2}$.

There are several points of interests for the two-point function on the $\eta$-plane. The first is the limit $\eta=0$. Physically, it means that the two operators coincide in Euclidean spacetime (or light-like separated in Lorentzian spacetime). In the limit of coinciding operators, we can use the standard OPE 
\begin{equation}\label{bulkOPE}
\mathcal{O}_{\Delta_1}(x_1)\mathcal{O}_{\Delta_2}(x_2)=\frac{\delta_{\Delta_1\Delta_2}}{(x_1-x_2)^{2\Delta_1}}+\sum_kC_{12k} D[x_1-x_2,\partial_{x_2}]\mathcal{O}_{\Delta_k}(x_2)
\end{equation}
where $k$ labels the conformal primaries, and $C_{12k}$ is the OPE coefficient. The differential operator $D[x_1-x_2,\partial_{x_2}]$ is completely determined by the conformal symmetry. The OPE reduces the two-point function to one-point functions which are completely determined by kinematics up to an overall constant. The contribution of each primary operator in the OPE to the two-point function can be resummed into a conformal block \cite{Nakayama:2016xvw}
\begin{equation}\label{defg}
g_\Delta(\eta)=\eta^{\frac{\Delta-\Delta_1-\Delta_2}{2}} {}_2F_1\left(\frac{\Delta+\Delta_1-\Delta_2}{2},\frac{\Delta+\Delta_2-\Delta_1}{2};\Delta-\frac{d}{2}+1;\eta\right)\;.
\end{equation}
The conformal block can also be obtained more conveniently from solving the conformal Casimir equation
\begin{equation}
L^2 \langle \mathcal{O}_{\Delta_1}(P_1)\mathcal{O}_{\Delta_2}(P_2)\rangle=-\Delta(\Delta-d)\langle \mathcal{O}_{\Delta_1}(P_1)\mathcal{O}_{\Delta_2}(P_2)\rangle\;,
\end{equation}
with the boundary condition 
\begin{equation}\label{gbc}
g_{\Delta}(\eta)\sim \eta^{\frac{\Delta-\Delta_1-\Delta_2}{2}}\;,\quad \eta\to0\;.
\end{equation}
The Casimir operator 
\begin{equation}\label{defL2}
L^2=\frac{1}{2}(L_1^{AB}+L_2^{AB})(L_{1,AB}+L_{2,AB})
\end{equation}
is constructed from the $SO(d,2)$ generators
\begin{equation}
L_{AB}=P_A\frac{\partial}{\partial P^B}-P_B\frac{\partial}{\partial P^A}\;.
\end{equation}
In terms of these conformal blocks, the two-point function can be written as 
\begin{equation}
\mathcal{G}(\eta)=\sum_{k} \mu_{12k} g_{\Delta_k}(\eta)
\end{equation}
where 
\begin{equation}
\mu_{12k}=a_kC_{12k}\;.
\end{equation}

Similarly, $\eta=1$ is also an interesting point where one operator approaches the image of the other operator (or the lightcone of the image). We can again apply the OPE for an operator with an image operator, which is equivalent to the original OPE thanks to (\ref{opidentification}). We will refer to this channel as the {\it image channel}. The two-point function can be decomposed into the {\it image conformal blocks}
\begin{equation}\label{defgbar}
\bar{g}_\Delta(\eta)=(1-\eta)^{\frac{\Delta-\Delta_1-\Delta_2}{2}} {}_2F_1\left(\frac{\Delta+\Delta_1-\Delta_2}{2},\frac{\Delta+\Delta_2-\Delta_1}{2};\Delta-\frac{d}{2}+1;1-\eta\right)\;.
\end{equation}
These image conformal blocks are eigenfunctions of the image conformal Casimir equation
\begin{equation}
\bar{L}^2 \langle \mathcal{O}_{\Delta_1}(P_1)\mathcal{O}_{\Delta_2}(P_2)\rangle=-\Delta(\Delta-d)\langle \mathcal{O}_{\Delta_1}(P_1)\mathcal{O}_{\Delta_2}(P_2)\rangle\;,
\end{equation}
with the boundary condition 
\begin{equation}\label{gbarbc}
\bar{g}_{\Delta}(\eta)\sim (1-\eta)^{\frac{\Delta-\Delta_1-\Delta_2}{2}}\;,\quad \eta\to1\;.
\end{equation}
Here the image conformal Casimir operator 
\begin{equation}\label{defLbar2}
\bar{L}^2=\frac{1}{2}(L_1^{AB}+\bar{L}_2^{AB})(L_{1,AB}+\bar{L}_{2,AB})
\end{equation}
involves the generators at the image position
\begin{equation}
\bar{L}_{AB}=\bar{P}_A\frac{\partial}{\partial \bar{P}^B}-\bar{P}_B\frac{\partial}{\partial \bar{P}^A}
\end{equation}
where $\bar{P}^A$ is the embedding vector for the image point
\begin{equation}
\bar{P}^A=\left(\frac{1+x^{-2}}{2},\frac{1-x^{-2}}{2},-\frac{x^\mu}{x^2}\right)\;.
\end{equation}
In terms of the conformal blocks, the crossing equation (\ref{crossingeqn}) now reads
\begin{equation}
\sum_{k} \mu_{12k} (g_{\Delta_k}(\eta)\mp \bar{g}_{\Delta_k}(\eta))=0\;.
\end{equation}

Finally, there is another interesting point $\eta=\infty$, which can only be reached in Lorentzian signature. It turns out, as we will see in Section \ref{Sec:3}, that this limit plays a similar role as the ``Regge limit'' in BCFT two-point functions. In fact, the kinematics of boundary CFTs are intimately related to $\mathbb{RP}^d$ CFTs. We now give a detailed comparison with the closely related BCFT case,\footnote{In the table and discussion below, we take the conformal boundary of the BCFT to be a unit sphere. It can be mapped to the infinite plane boundary by a conformal transformation.} and the result is summarized in Table \ref{table1}.
{\begin{table}
\begin{center}
\begin{tabular}{||c|| c | c ||} 
 \hline\hline
& $\mathbb{RP}^d$ CFT & BCFT$_d$ \\ [0.5ex] 
 \hline\hline
One-point function & $\langle \mathcal{O}_\Delta(x)\rangle= \frac{a_{\mathcal{O}}}{(1+x^2)^\Delta}$ &  $\langle \mathcal{O}_\Delta(x)\rangle_B=\frac{a_{B,\mathcal{O}}}{|1-x^2|^\Delta}$\\ [0.5ex] 
 \hline
Two-point cross ratio & $\eta=\frac{(x_1-x_2)^2}{(1+x_1^2)(1+x_2^2)}$  & $\xi=\frac{(x_1-x_2)^2}{(1-x_1^2)(1-x_2^2)}$ \\ [0.5ex] 
\hline
Two-point function & $G=\frac{1}{(1+x_1^2)^{\Delta_1}(1+x_2^2)^{\Delta_2}}\mathcal{G}(\eta)$ &  $G_B=\frac{1}{|1-x_1^2|^{\Delta_1}|1-x_2^2|^{\Delta_2}} \mathcal{G}_B(\xi)$\\ [0.5ex] 
\hline
OPE limits & $\eta\to 0$ (bulk channel) & $\xi\to 0$ (bulk channel)\\
& $\eta \to 1$ (image channel) &  $\xi\to\infty$ (boundary channel)\\ [0.5ex] 
\hline
Regge limit& $\eta\to\infty$& $\xi\to-1$\\ [0.5ex] 
\hline\hline
\end{tabular}
\caption{A comparison of kinematics for $\mathbb{RP}^d$ CFTs and boundary CFTs.}
\label{table1}
\end{center}
\end{table}}

\subsubsection*{Intermezzo: comparing with boundary CFTs}
As we mentioned in the last section, a boundary CFT preserves the conformal symmetry that leaves $N_B$ invariant. Up to a Wick rotation, the two systems preserve the same symmetry group. The inversion sphere $x^2=1$ now becomes a fixed locus in the BCFT case, and is the location of the conformal boundary. The one-point function of an operator inserted away from the boundary is determined by kinematics\footnote{In order to distinguish from the real projective space case, we use the subscript $B$ to denote objects in boundary CFTs.}
\begin{equation}\label{B1pt}
\langle\mathcal{O}_\Delta(x)\rangle_B=\frac{a_{B,\mathcal{O}}}{|2P\cdot N_B|^\Delta}=\frac{a_{B,\mathcal{O}}}{|1-x^2|^\Delta}\;.
\end{equation}
up to a constant $a_{B,\mathcal{O}}$. The one-point coefficients $a_{B,\mathcal{O}}$ are part of the new data defining a BCFT. For two operators inserted away from the boundary\footnote{We will assume that the operators are inserted on the same side of the boundary.}, we can construct a cross ratio
\begin{equation}\label{defxi}
\xi=\frac{(-2 P_1\cdot P_2)}{(2N_B\cdot P_1)(2N_B\cdot P_2)}=\frac{(x_1-x_2)^2}{(1-x_1^2)(1-x_2^2)}\;,
\end{equation}
which is invariant under the residual conformal symmetry and the independent rescaling of the operators. In a Euclidean spacetime, the range of the cross ratio is $\xi\in[0,\infty)$. The two-point function can be written as a function of the cross ratio
\begin{equation}
\langle \mathcal{O}_{\Delta_1}(x_1)\mathcal{O}_{\Delta_1}(x_2)\rangle_B=\frac{\mathcal{G}_B(\xi)}{|2N_B\cdot P_1|^{\Delta_1}|2N_B\cdot P_2|^{\Delta_1}}=\frac{\mathcal{G}_B(\xi)}{|1-x_1^2|^{\Delta_1}|1-x_2^2|^{\Delta_2}}\;.
\end{equation}

There are three interesting points on the $\xi$-plane. The point $\xi=0$ is known as the bulk channel OPE limit, and should be identified with the $\eta=0$ case where operators coincide (or light-like separated in Lorentzian signature). We can apply the OPE (\ref{bulkOPE}), and reduce the two-point function into one-point functions. The two-point function can be written as a sum of bulk channel conformal blocks \cite{McAvity:1995zd,Liendo:2012hy}
\begin{equation}
\mathcal{G}_B(\xi)=\sum_k \mu_{B,12k} g^{bulk}_{B,\Delta_k}(\xi)
\end{equation}
where $\mu_{B,12k}=C_{12k}a_{B,k}$ and 
\begin{equation}\label{defgBbulk}
g^{bulk}_{B,\Delta}(\xi)=\xi^{\frac{\Delta-\Delta_1-\Delta_2}{2}} {}_2F_1\left(\frac{\Delta+\Delta_1-\Delta_2}{2},\frac{\Delta+\Delta_2-\Delta_1}{2};\Delta-\frac{d}{2}+1;-\xi\right)\;.
\end{equation}
Note that $g^{bulk}_{B,\Delta}(\xi)$ can be identified with $g_\Delta(\eta)$ with the replacement $\xi\leftrightarrow -\eta$,\footnote{There is some formal connection between boundary CFTs and $\mathbb{RP}^d$ CFTs by analytic continuation. For example, in two dimensions the boundary states are defined by $(L_n-\bar{L}_{-n})|\mathcal{B}\rangle=0$ while the crosscap states are defined by $(L_n-(-1)^n\bar{L}_{-n})|\mathcal{C}\rangle=0$. Here we can relate them formally by making the analytic continuation $x^2\to -x^2$. This not only gives $\xi\to-\eta$, but also fixes the one-point function and the kinematic factors we extracted from the two-point functions.} up to an overall normalization.The limit $\xi=\infty$ is known as the boundary channel limit, where operators inserted in the bulk are taken close to the boundary. A different OPE is involved in this limit 
\begin{equation}
\mathcal{O}_\Delta(x)=\frac{a_{B,\mathcal{O}}}{|1-x^2|^\Delta}+\sum_l \rho_\ell C[x]\widehat{O}_{\Delta_l}(x)
\end{equation}
which expresses a bulk operator as a sum of operators $\widehat{O}_{\Delta_l}(x)$ on the boundary. Here $C[x]$ is a differential operator determined by symmetry. The boundary operator spectrum $\Delta_l$ and OPE coefficient $\rho_l$ are new CFT data. Applying the boundary OPE for each operator reduces the two-point function to a sum of two-point functions of operators living on the boundary, and the latter is fully fixed by the residual conformal symmetry. The contribution of exchanging a boundary operator is summarized by a boundary channel conformal block 
\begin{equation}
g^{boundary}_{B,\Delta}(\xi)=\xi^{-\Delta}{}_2F_1\left(\Delta,\Delta-\frac{d}{2}+1;2\Delta+2-d,-\frac{1}{\xi}\right)\;,
\end{equation}
and the correlator can be decomposed in the boundary channel as 
\begin{equation}
\mathcal{G}_B(\xi)=a_{B,\mathcal{O}}^2\delta_{12}+\sum_l \rho_{1,l}\rho_{2,l}g^{boundary}_{B,\Delta_l}(\xi)\;.
\end{equation}
The boundary channel of BCFT two-point functions has no analogue in the real projective space case, because the identification (\ref{inversion}) does not have any fixed point. The two channels of OPE should lead to the same answer, and gives to a ``crossing equation''
\begin{equation}
\sum_k \mu_{B,12k} g^{bulk}_{B,\Delta_k}(\xi)=a_{B,\mathcal{O}}^2\delta_{12}+\sum_l \rho_{1,l}\rho_{2,l}g^{boundary}_{B,\Delta_l}(\xi)\;.
\end{equation}
Finally, the limit of $\xi=-1$ is known as the ``Regge limit'' \cite{Mazac:2018biw}.
In this limit, one operator is at the lightcone created by the image of the other operator with respect to the boundary.  The Regge limit can only be reached in the Lorentzian signature, and requires analytic continuation from the Euclidean signature. It was proven in \cite{Mazac:2018biw} that for any unitary boundary CFT, the two-point function has a bounded behavior at the Regge limit, which is controlled by the bulk channel exchange of operators with the lowest dimension.

We can think of the $\eta=\infty$ limit of $\mathbb{RP}^d$ CFTs as the $\xi=-1$ limit of BCFTs, as both cases requires analytic continuation from the Euclidean regime.\footnote{It might also be tempting to identify $\eta=1$ with the BCFT Regge limit, since in both cases one operator approaches the lightcone of (or coincides with) the image of the other operator. However, the crossing equation (\ref{crossingeqn}) tells us that $\eta=1$ limit is physically not any different from the $\eta=0$ limit.} This intuition will also be supported in the next section when we study Witten diagrams, which arise in the weakly coupled duals of $\mathbb{RP}^d$ CFTs.

\section{Holography on quotient AdS and Witten diagrams}\label{Sec:3}
\subsection{$AdS_{d+1}/\mathbb{Z}_2$}
In this section, we study a simple toy model of holography for $\mathbb{RP}^d$ CFTs. We extend the quotient of the boundary spacetime into the bulk to define a $\mathbb{Z}_2$ quotient of AdS space, and consider perturbative physics on this background. This over-simplified setup is effective in nature, and does not correspond to top-down models. However, it captures all the essential kinematics which are relevant to various applications later in the paper. This setup of quotient of AdS appeared previously, {\it e.g.}, in \cite{Verlinde:2015qfa, Maloney:2016gsg, Nakayama:2015mva,Nakayama:2016xvw,Lewkowycz:2016ukf}. Here we give a detailed account using the embedding space formalism introduced in Section \ref{Sec:2.1}.

\begin{figure}[htbp]
\begin{center}
\includegraphics[width=0.8\textwidth]{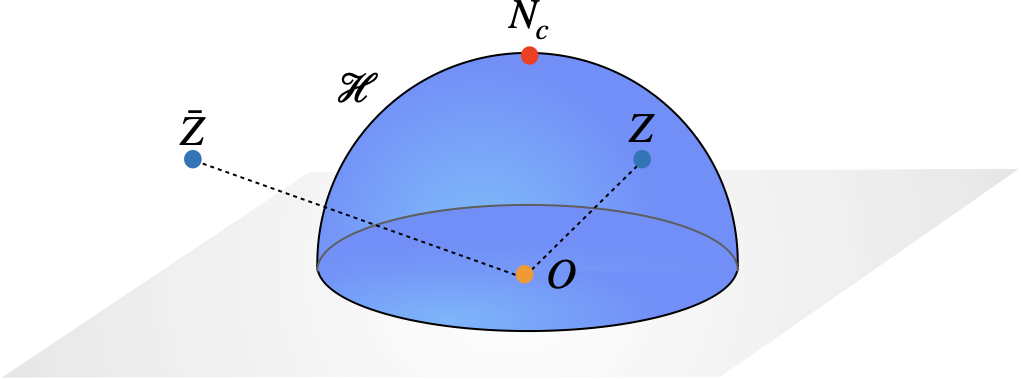}
\caption{Illustration of $AdS_{d+1}/\mathbb{Z}_2$ in Poincar\'e coordinates. A point $Z$ inside the hemisphere $\mathcal{H}$ is identified with its inversion image $\bar{Z}$ out side of the hemisphere. The quotient has a fixed point $N_c$, which is the north pole of the hemisphere.}
\label{Fig:AdSquotient}
\end{center}
\end{figure}

For the calculations in this section, it will be most convenient to consider the Euclidean AdS space\footnote{We have set the curvature of AdS to 1.} 
\begin{equation}
-\left(Z^1\right)^2+\left(Z^2\right)^2+\ldots+\left(Z^{d+2}\right)^2=-1\;, Z^1>0\;,
\end{equation}
and analytic continue the results to the Lorentzian signature in the end. In terms of the Poincar\'e coordinates $z=(z_0,\vec{z})$, the embedding space vector $Z^A$ is parameterized as 
\begin{equation}
Z^A=\frac{1}{z_0}\left(\frac{1+z_0^2+\vec{z}^2}{2},\frac{1-z_0^2-\vec{z}^2}{2},\vec{z}\right)\;.
\end{equation}
We extend the boundary inversion (\ref{calIembed}) into the bulk by requiring that $\mathcal{I}$ should act in the same way on the AdS embedding vector. This leads to
\begin{equation}\label{AdScalI}
\mathcal{I}:\quad z_0\to \frac{z_0}{z_0^2+\vec{z}^2}\;,\quad \vec{z}\to-\frac{\vec{z}}{z_0^2+\vec{z}^2}\;,
\end{equation}
and defines a quotient space $qAdS_{d+1}\equiv AdS_{d+1}/\mathbb{Z}_2$ by the identification 
\begin{equation}\label{AdSquotient}
z_0\leftrightarrow \frac{z_0}{z_0^2+\vec{z}^2}\;,\quad \vec{z}\leftrightarrow-\frac{\vec{z}}{z_0^2+\vec{z}^2}\;.
\end{equation}
Note that the identification (\ref{AdSquotient}) is geometrically represented in Poincar\'e coordinates by an inversion with respect to the hemisphere $\mathcal{H}$ defined by 
\begin{equation}\label{defcalH}
z_0^2+z^2=1\;,\quad z_0\geq 0\;,
\end{equation}
This is illustrated by Figure \ref{Fig:AdSquotient}. The map (\ref{AdSquotient}) has a fixed point
\begin{equation}
z_0=1\;,\quad \vec{z}=0\;,
\end{equation}
which sits at the north pole of the inversion hemisphere $\mathcal{H}$. In terms of embedding coordinates, the fixed point corresponds to the fixed vector $N_c$ introduced in (\ref{Nc}).

Now we consider a scalar field $\varphi_\pm$ living on the quotient space $qAdS_{d+1}$. To describe the scalar field, it is convenient to extend the definition of the scalar field to the full $AdS_{d+1}$ and impose the condition that 
\begin{equation}
\varphi_\pm(z)=\pm\, \varphi_\pm(\mathcal{I}\circ z)\;.
\end{equation}
We can define the propagators of $\varphi_\pm$ on $qAdS_{d+1}$ (and extended to $AdS_{d+1}$) as follows. The bulk-to-bulk propagator $H_{BB}^{\Delta,\pm}(Z,W)$ satisfies the following equation of motion 
\begin{equation}
\left(\square_Z+\Delta(\Delta-d)\right)H_{BB}^{\Delta,\pm}(Z,W)=\delta^{d+1}(Z,W)\pm\delta^{d+1}(Z,\bar{W})\;,
\end{equation}
where $\bar{W}$ denotes the image point of $W$ under the inversion (\ref{AdScalI}). The propagator $H_{BB}^{\Delta,\pm}(Z,W)$ can be expressed in terms of the usual AdS propagators before taking the quotient, as
\begin{equation}\label{HBB}
H_{BB}^{\Delta,\pm}(Z,W)=G^\Delta_{BB}(Z,W)\pm G^\Delta_{BB}(Z,\bar{W})
\end{equation}
where $G^\Delta_{BB}(Z,W)$ satisfies 
\begin{equation}\label{eomGBB}
\left(\square_Z+\Delta(\Delta-d)\right)G_{BB}^{\Delta}(Z,W)=\delta^{d+1}(Z,W)\;,
\end{equation}
and is given explicitly by
\begin{equation}
G^\Delta_{BB}(Z,W)=C^{\Delta,d}_{BB}(-u)^{-\Delta}{}_2F_1\left(\Delta,\Delta-\frac{d}{2}+\frac{1}{2};2\Delta-d+1;u^{-1}\right)\;,
\end{equation}
with
\begin{equation}
u=-\frac{Z\cdot W+1}{2}\;,\quad C^{\Delta,d}_{BB}=\frac{\Gamma(\Delta)\Gamma(\Delta-\frac{d}{2}+\frac{1}{2})}{(4\pi)^{\frac{d+1}{2}}\Gamma(2\Delta-d+1)}\;.
\end{equation}
Note that 
\begin{equation}
Z\cdot W=\bar{Z}\cdot \bar{W}\;,\quad Z\cdot \bar{W}=\bar{Z}\cdot W\;,
\end{equation}
we have the following identities for the bulk-to-bulk propagator
\begin{equation}\label{HBBidentities}
H_{BB}^{\Delta,\pm}(Z,W)=H_{BB}^{\Delta,\pm}(W,Z)=H_{BB}^{\Delta,\pm}(\bar{Z},\bar{W})\;, \quad H_{BB}^{\Delta,\pm}(Z,\bar{W})=\pm H_{BB}^{\Delta,\pm}(Z,W)\;.
\end{equation}
The bulk-to-boundary propagator $H_{B\partial}^{\Delta,\pm}(Z,P)$ can be obtained from the limit of $H_{BB}^{\Delta,\pm}(Z,W)$ as we move the bulk point $W$ close to the boundary. It is easy to prove that  
\begin{equation}
Z\cdot \bar{P}=(\vec{x}^2)^{-1}\bar{Z}\cdot P\;,\quad \bar{Z}\cdot \bar{P}=(\vec{x}^2)^{-1}Z\cdot P\;.
\end{equation}
This implies that the usual $AdS_{d+1}$ bulk-to-boundary propagator  
\begin{equation}
G^\Delta_{B\partial}(Z,P)=\left(\frac{1}{-2Z\cdot P}\right)^\Delta=\left(\frac{z_0}{z_0^2+(\vec{z}-\vec{x})^2}\right)^\Delta
\end{equation}
transforms as 
\begin{equation}\label{GBpartialinversion}
G^\Delta_{B\partial}(Z,\bar{P})=(\vec{x}^2)^\Delta G^\Delta_{B\partial}(\bar{Z},P)\;.
\end{equation}
On the other hand, we recall that boundary operator receives an extra $(\vec{x}^2)^{-\Delta}$ under inversion from (\ref{opidentification}), which cancels the $(\vec{x}^2)^{\Delta}$ in (\ref{GBpartialinversion}). Therefore the $qAdS_{d+1}$ bulk-to-boundary propagator should be defined similarly to (\ref{HBB}), as 
\begin{equation}\label{HBpartiala}
H_{B\partial}^{\Delta,\pm}(Z,P)=G^\Delta_{B\partial}(Z,P)\pm G^\Delta_{B\partial}(\bar{Z},P)\;,
\end{equation}
or equivalently, 
\begin{equation}\label{HBpartialb}
H_{B\partial}^{\Delta,\pm}(Z,P)=G^\Delta_{B\partial}(Z,P)\pm (\vec{x}^2)^{-\Delta}G^\Delta_{B\partial}(Z,\bar{P})\;.
\end{equation}

\subsection{Tree-level Witten diagrams on quotient AdS space}\label{Sec:3.2}
Having obtained bulk-to-bulk and bulk-to-boundary propagators on $qAdS_{d+1}$, we are now ready to define Witten diagrams. 

\vspace{0.5cm}
\noindent{\bf One-point diagram} 
\vspace{0.1cm}

\begin{figure}[htbp]
\begin{center}
\includegraphics[width=0.8\textwidth]{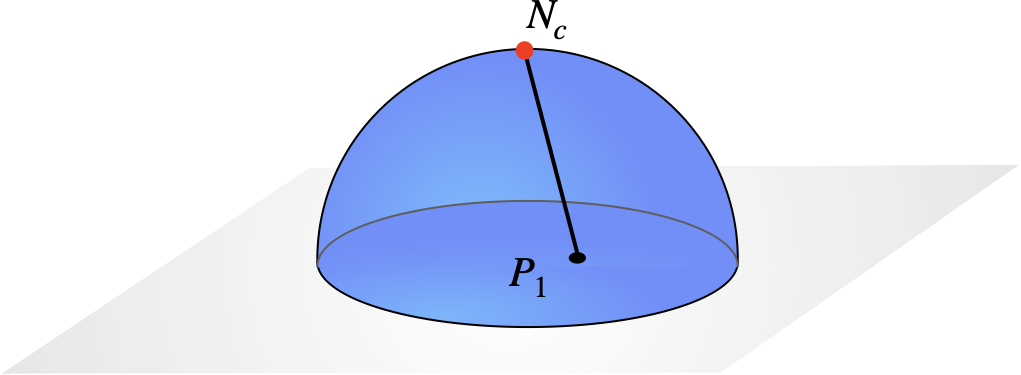}
\caption{The one-point function is given by a bulk-to-boundary propagator with no integration.}
\label{Fig:1pt}
\end{center}
\end{figure}

\noindent  Let us start from the one-point function. It is given by a single bulk-to-boundary propagator which ends on the conformal boundary where the operator is inserted. Note that the Witten diagram must preserve the fixed vector $N_c$ defined in (\ref{Nc}). Therefore, the other end of the propagator has to end at the inversion fixed point $N_c$ (Figure \ref{Fig:1pt}). It corresponds to a vertex $\varphi(N_c)$ localized at the fixed point. We have
\begin{equation}\label{AdS1pt}
\langle \mathcal{O}_\Delta \rangle=H_{B\partial}^{\Delta,\pm}(N_c,P)=\left\{\begin{array}{l}\frac{2}{(1+\vec{x}^2)^\Delta}\quad \text{for }+ \text{ parity}\\ 0\quad\quad\quad\;\;\, \text{for }- \text{ parity}\end{array}\right.
\end{equation} 
which has the correct structure (\ref{1ptfun}) determined by symmetry. Note that the information of the vertex $\varphi(N_c)$ was not contained in the original theory before the quotient, but was inputted into this toy model by hand.

\vspace{0.5cm}
\noindent{\bf Two-point contact Witten diagrams} 
\vspace{0.1cm}

\begin{figure}[htbp]
\begin{center}
\includegraphics[width=0.8\textwidth]{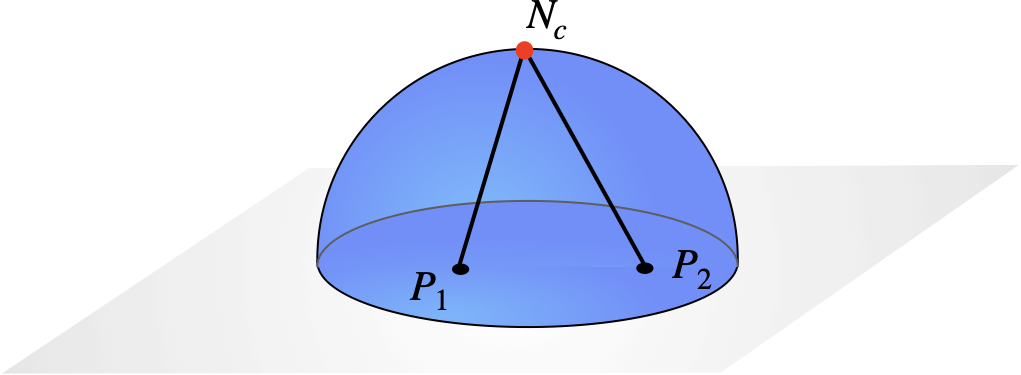}
\caption{The two-point contact Witten diagram is given by a product of bulk-to-boundary propagators with no integration.}
\label{Fig:con}
\end{center}
\end{figure}

\noindent For two-point functions, we can define the following tree-level contact Witten diagram (Figure \ref{Fig:con})
\begin{equation}\label{Vcon0}
V^{con,0}_{\Delta_1,\Delta_2}(P_1,P_2)=\frac{1}{4}H_{B\partial}^{\Delta_1,+}(N_c,P_1)H_{B\partial}^{\Delta_2,+}(N_c,P_2)\;,
\end{equation}
which factorizes into the product of two one-point functions. It comes from the vertex $\frac{1}{4}\varphi_1\varphi_2(N_c)$ which localizes at the fixed point $N_c$. Note that it is important that both operators are parity even since $H_{B\partial}^{\Delta,-}(N_c,P_1)$ vanishes. In terms of the cross ratio, the contact Witten diagram reads  
\begin{equation}\label{defV0a}
V^{con,0}_{\Delta_1,\Delta_2}(P_1,P_2)=\frac{\mathcal{V}^{con,0}_{\Delta_1,\Delta_2}(\eta)}{(1+x_1^2)^{\Delta_1}(1+x_2^2)^{\Delta_2}}
\end{equation}
where 
\begin{equation}\label{defV0b}
\mathcal{V}^{con,0}_{\Delta_1,\Delta_2}(\eta)=1\;.
\end{equation}

To define two-point contact diagram for operators with odd parity, we can add two derivatives to the vertex. The vertex now becomes $\frac{1}{4}(\nabla^\mu \varphi_1 \nabla_\mu \varphi_2)(N_c)$, and leads to the following diagram 
\begin{equation}
\begin{split}
V^{con,2}_{\Delta_1,\Delta_2}(P_1,P_2)={}&\frac{1}{4}\nabla^\mu H_{B\partial}^{\Delta_1,-}(Z,P_1)\nabla_\mu H_{B\partial}^{\Delta_2,-}(Z,P_2)\big|_{Z=N_c}\;,\\
={}&\nabla^\mu G_{B\partial}^{\Delta_1}(Z,P_1)\nabla_\mu G_{B\partial}^{\Delta_2}(Z,P_2)\big|_{Z=N_c}\;.
\end{split}
\end{equation}
Using the identity 
\begin{equation}
\begin{split}
\nabla_\mu G^{\Delta_1}_{B\partial}(Z,P_1) \nabla^\mu G^{\Delta_2}_{B\partial}(Z,P_2)={}&\Delta_1\Delta_2\bigg(G^{\Delta_1}_{B\partial}(Z,P_1)G^{\Delta_2}_{B\partial}(Z,P_2)\\
{}&-2x_{12}^2G^{\Delta_1+1}_{B\partial}(Z,P_1)G^{\Delta_2+1}_{B\partial}(Z,P_2)\bigg)\;,
\end{split}
\end{equation}
we get 
\begin{equation}
V^{con,2}_{\Delta_1,\Delta_2}(P_1,P_2)=\frac{\mathcal{V}^{con,2}_{\Delta_1,\Delta_2}(\eta)}{(1+x_1^2)^{\Delta_1}(1+x_2^2)^{\Delta_2}}
\end{equation}
where 
\begin{equation}
\mathcal{V}^{con,2}_{\Delta_1,\Delta_2}(\eta)=\Delta_1\Delta_2(1-2\eta)\;.
\end{equation}
It is clear that $\mathcal{V}^{con,2}_{\Delta_1,\Delta_2}(\eta)$ is antisymmetric under $\eta\leftrightarrow 1-\eta$.

It is straightforward to generalize these contact Witten diagrams to include more derivatives in the vertex. In general, for a vertex with $2L$ derivatives the two-point contact Witten diagram $\mathcal{V}^{con,2L}_{\Delta_1,\Delta_2}(\eta)$ is a polynomial in $\eta$ of degree $L$. Moreover, the contact Witten diagrams have the following behavior at $\eta\to\infty$
\begin{equation}\label{VconRegge}
\mathcal{V}^{con,2L}_{\Delta_1,\Delta_2}(\eta)\to \eta^L\;,\quad \eta\to \infty\;.
\end{equation}
This is consistent with the intuition that the $\eta\to\infty$ limit can be thought of as a ``Regge limit'' for the two-point correlator -- increasing  the number of derivatives in the vertex leads to a more divergent behavior in the correlator at large cross ratio.

\vspace{0.5cm}
\noindent{\bf Two-point exchange Witten diagrams} 
\vspace{0.1cm}

\begin{figure}[htbp]
\begin{center}
\includegraphics[width=0.8\textwidth]{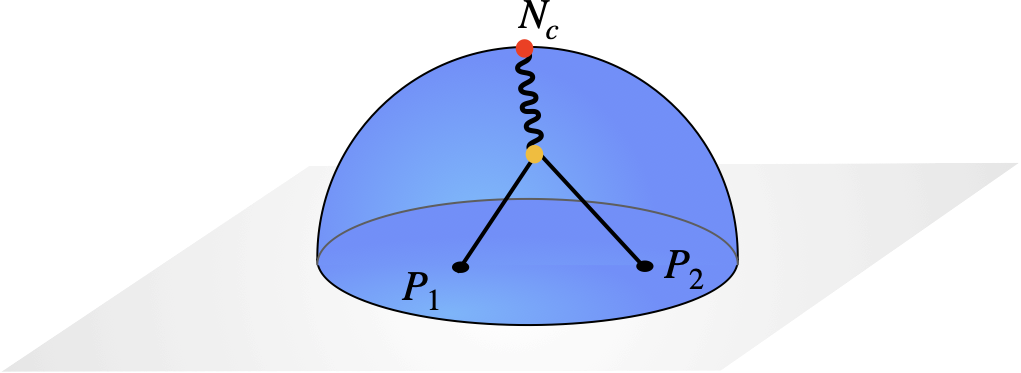}
\caption{Illustration of an exchange Witten diagram. One end of the bulk-to-bulk propagator is fixed at $N_c$, while the other end is connected to the cubic vertex and integrated over.}
\label{Fig:exch}
\end{center}
\end{figure}

\noindent Let us now define the two-point exchange Witten diagram (Figure \ref{Fig:exch})
\begin{equation}
V^{exchange,\pm}_\Delta(P_1,P_2)=\frac{1}{2}\int_{qAdS_{d+1}}d^{d+1}Z\; H^{\Delta,+}_{BB}(N_c,Z)H_{B\partial}^{\Delta_1,\pm}(Z,P_1)H_{B\partial}^{\Delta_2,\pm}(Z,P_2)
\end{equation}
which requires the presence of the localized vertex $\varphi(N_c)$ for the scalar field with dual conformal dimension $\Delta$, and a bulk cubic vertex $\varphi \varphi_1\varphi_2(Z)$ where $\varphi_{1,2}$ have dual dimensions $\Delta_{1,2}$. Let us use (\ref{HBB}), (\ref{HBpartiala}) and (\ref{HBpartialb}) to express the $qAdS_{d+1}$ propagators in terms of the propagators in the full $AdS_{d+1}$. Note that $N_c$ is its own image, therefore $H^{\Delta,+}_{BB}(N_c,Z)=2G^\Delta_{BB}(Z,N_c)$. On the other hand, $H^{\Delta,+}_{BB}(N_c,Z)=2G^\Delta_{BB}(\bar{Z},N_c)$ thanks to (\ref{HBBidentities}). Using this and (\ref{GBpartialinversion}), we can expand the product and massage the expressions such that the $AdS_{d+1}$ propagators join into connected diagrams. It becomes obvious that the integrals in $V^{exchange,\pm}_\Delta(P_1,P_2)$ can be organized into the linear combinations of exchange diagrams defined in the full $AdS_{d+1}$ space (in particular, integrals inside the sphere and their images outside combine, and extend to the full space)
\begin{equation}\label{VasWWbar}
V^{exchange,\pm}_\Delta(P_1,P_2)=W^{exchange}_\Delta(P_1,P_2)\pm (x_2^2)^{-\Delta_2}\bar{W}^{exchange,\pm}_\Delta(P_1,P_2)
\end{equation}
where 
\begin{equation}\label{defW}
W^{exchange}_\Delta(P_1,P_2)=\int_{AdS_{d+1}}d^{d+1}Z\; G^{\Delta}_{BB}(N_c,Z)G_{B\partial}^{\Delta_1}(Z,P_1)G_{B\partial}^{\Delta_2}(Z,P_2)\;,
\end{equation}
and 
\begin{equation}\label{defWbar}
\bar{W}^{exchange}_\Delta(P_1,P_2)=W^{exchange}_\Delta(P_1,\bar{P}_2)\;.
\end{equation}
When written in terms of the cross ratio
\begin{equation}
\begin{split}
{}&W^{exchange}_\Delta(P_1,P_2)=\frac{\mathcal{W}^{exchange}_\Delta(\eta)}{(1+x_1^2)^{\Delta_1}(1+x_2^2)^{\Delta_2}}\;,\\ 
{}&(x_2^2)^{-\Delta_2}\bar{W}^{exchange}_\Delta(P_1,P_2)=\frac{\bar{\mathcal{W}}^{exchange}_\Delta(\eta)}{(1+x_1^2)^{\Delta_1}(1+x_2^2)^{\Delta_2}}\;,
\end{split}
\end{equation}
$\mathcal{W}^{exchange}_\Delta$ and $\bar{\mathcal{W}}^{exchange}_\Delta$ are related by
\begin{equation}\label{calWWbcrossing}
\bar{\mathcal{W}}^{exchange}_\Delta(\eta)=\mathcal{W}^{exchange}_\Delta(1-\eta)\;.
\end{equation}

\vspace{0.5cm}
\noindent{\bf Equation of motion relations} 
\vspace{0.1cm}

The exchange Witten diagrams $W^{exchange}_\Delta$ and $\bar{W}^{exchange}_\Delta$ introduced above are related to the the zero-derivative contact Witten diagram $V^{con,0}_{\Delta_1,\Delta_2}$ by the equation of motion operators. More precisely, let us define 
\begin{equation}
\begin{split}
\mathbf{EOM}={}& L^2+\Delta(\Delta-d)\;,\\
\overline{\mathbf{EOM}}={}& \bar{L}^2+\Delta(\Delta-d)
\end{split}
\end{equation}
where $L^2$ and $\bar{L}^2$ are the conformal Casimir operators defined in (\ref{defL2}) and (\ref{defLbar2}). The exchange diagrams are then related to the contact diagram via 
\begin{equation}
\begin{split}\label{eomWtoV}
\mathbf{EOM}[W^{exchange}_\Delta]={}&V^{con,0}_{\Delta_1,\Delta_2}\;,\\
\overline{\mathbf{EOM}}[\bar{W}^{exchange}_\Delta]={}&V^{con,0}_{\Delta_1,\Delta_2}\;.
\end{split}
\end{equation}
To prove these relations, let us notice that the integral (\ref{defW}) is conformal invariant  
\begin{equation}\label{covW}
(L_1^{AB}+L_2^{AB}+\mathcal{L}^{AB})W^{exchange}_\Delta(P_1,P_2;N_c)=0\;.
\end{equation}
Here we are viewing $W^{exchange}_\Delta$ as a function of the bulk point $N_c$, and $\mathcal{L}^{AB}$ are the $AdS_{d+1}$ isometry generators at the point $N_c$. Using (\ref{covW}) twice, we get 
\begin{equation}
L^2W^{exchange}_\Delta(P_1,P_2;N_c)=\mathcal{L}^2W^{exchange}_\Delta(P_1,P_2;N_c)
\end{equation}
where $\mathcal{L}^2=\frac{1}{2}\mathcal{L}^{AB}\mathcal{L}_{AB}$. Notice that $\mathcal{L}^2$ acts on the bulk coordinates as $\square$, and collapses the bulk-to-bulk propagator in (\ref{defW}) to a delta function by (\ref{eomGBB}). Integrating over $Z$ gives us the contact diagram (\ref{Vcon0}). The proof for $\bar{W}^{exchange}_\Delta$ is analogous. 

We will also explicitly evaluate these exchange Witten diagrams, and study their decompositions into conformal blocks. We will delay the discussion until Section \ref{Sec:4}.

\subsection{Geodesic Witten diagrams}\label{Sec:geoW}
\begin{figure}[htbp]
\begin{center}
\includegraphics[width=0.8\textwidth]{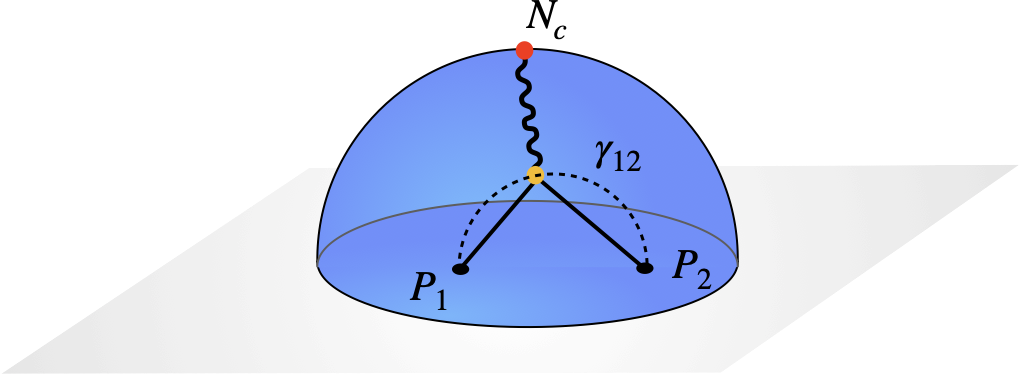}
\caption{Illustration of a geodesic Witten diagram. The integration of the cubic vertex is restricted to the geodesic line $\gamma_{12}$ connecting the two boundary insertions.}
\label{Fig:Wgeo}
\end{center}
\end{figure}

\begin{figure}[htbp]
\begin{center}
\includegraphics[width=0.8\textwidth]{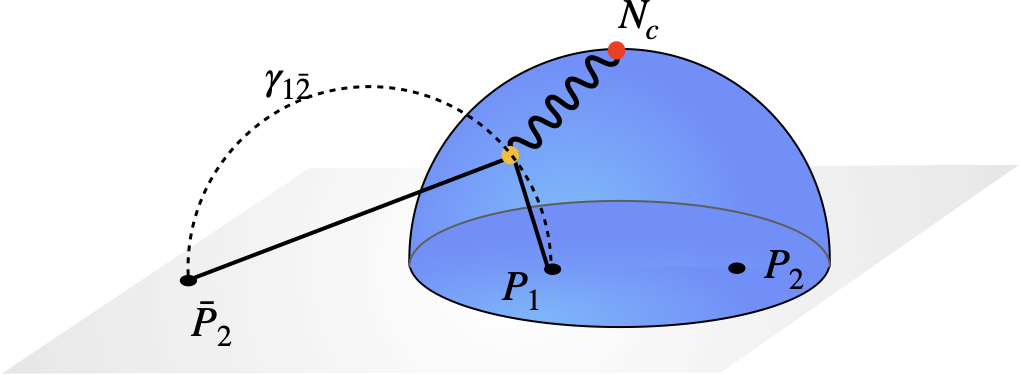}
\caption{Illustration of a geodesic Witten diagram in the mirror channel. The geodesic $\gamma_{1\bar{2}}$ now connects point 1 and the image of point 2.}
\label{Fig:Wgeob}
\end{center}
\end{figure}

We can also define a variation of the exchange Witten diagrams, which is holographically dual to the conformal blocks (\ref{defg}), (\ref{defgbar}).  These modified exchange Witten diagrams are known as the geodesic Witten diagrams \cite{Hijano:2015zsa}. These objects were also considered in \cite{daCunha:2016crm} in a different context. We define the dual of $g_\Delta(\eta)$ similarly as in (\ref{defW}) by 
\begin{equation}\label{defWgeo}
W^{geo}_\Delta(P_1,P_2)=\int_{\gamma_{12}}d\gamma \; G^{\Delta}_{BB}(N_c,\gamma)G_{B\partial}^{\Delta_1}(\gamma,P_1)G_{B\partial}^{\Delta_2}(\gamma,P_2)\;,
\end{equation}
where $\gamma_{12}$ is the geodesic connecting the $\vec{x}_1$ and $\vec{x}_2$ on the conformal boundary. In Poincar\'e coordinates, the geodesic $\gamma_{12}$ is just a semicircle.
Instead of integrating over the whole $AdS_{d+1}$, the integration is now restricted to the geodesic only. The geodesic Witten diagram is illustrated in Figure \ref{Fig:Wgeo}. After extracting a kinematic factor 
\begin{equation}
W^{geo}_\Delta(P_1,P_2)=\frac{\mathcal{W}^{geo}_\Delta(\eta)}{(1+x_1^2)^{\Delta_1}(1+x_2^2)^{\Delta_2}}\;,
\end{equation}
the function $\mathcal{W}^{geo}_\Delta(\eta)$ is proportional to $g_\Delta(\eta)$ up to some constant factor. Similarly, we define 
\begin{equation}\label{defWbargeo}
\bar{W}^{geo}_\Delta(P_1,P_2)=\int_{\gamma_{1\bar{2}}}d\gamma \; G^{\Delta}_{BB}(N_c,\gamma)G_{B\partial}^{\Delta_1}(\gamma,P_1)G_{B\partial}^{\Delta_2}(\gamma,\bar{P}_2)\;,
\end{equation}
and 
\begin{equation}
(x_2^2)^{-\Delta_2}\bar{W}^{geo}_\Delta(P_1,P_2)=\frac{\bar{\mathcal{W}}^{geo}_\Delta(\eta)}{(1+x_1^2)^{\Delta_1}(1+x_2^2)^{\Delta_2}}\;,
\end{equation}
where $\gamma_{1\bar{2}}$ is the geodesic connecting $\vec{x}_1$ and the image of $\vec{x}_2$ (see Figure \ref{Fig:Wgeob}). The geodesic Witten diagram $\bar{W}^{geo}_\Delta$ is dual to the image conformal block $\bar{g}_\Delta(\eta)$.

To prove their equivalence, we can use the equation of motion operators introduced in the last subsection. Let us act on the geodesic Witten diagrams with $\mathbf{EOM}$ and $\overline{\mathbf{EOM}}$, and use the definitions of $W^{geo}_\Delta$ and $\bar{W}^{geo}_\Delta$. The integrations along the geodesics preserve the conformal invariance, and allows us to apply the analysis and use the equation of motion for the bulk-to-bulk propagator. Note however that since the geodesic lines do not pass through the fixed point $N_c$ for generic end points $\vec{x}_1$, $\vec{x}_2$, the delta function is not integrated. Therefore, instead of generating contact diagrams we get 
\begin{equation}
\mathbf{EOM}[W^{geo}_\Delta]=0\;, \quad \overline{\mathbf{EOM}}[\bar{W}^{geo}_\Delta]=0\;,
\end{equation}
On the boundary side, these two equations are just the quadratic conformal Casimir equations. It is also clear from the definitions (\ref{defWgeo}), $(\ref{defWbargeo}$) that $\mathcal{W}^{geo}_\Delta(\eta)$ and $\bar{\mathcal{W}}^{geo}_\Delta(\eta)$ satisfy the boundary conditions (\ref{gbc}), (\ref{gbarbc}). This concludes the proof that the geodesic Witten diagrams are the bulk dual of the conformal blocks.

\subsection{Comparison with interface CFT from the probe-brane setup}
To conclude this section, we give a quick comparison of the above discussion with the closely related interface/boundary CFT from the probe-brane setup \cite{DeWolfe:2001pq,Aharony:2003qf,Rastelli:2017ecj,Mazac:2018biw} (which is the simplest Karch-Randall set up \cite{Karch:2000gx,Karch:2001cw}). As we have seen in Section \ref{Sec:2}, the kinematics of the $\mathbb{RP}^d$  CFT and BCFT share a lot of similarities. We now show how these kinematical similarities are extended into the bulk, and also point out a number of differences. 

In the probe-brane setup, we choose a special slice of $AdS_d$ space inside $AdS_{d+1}$. There are local degrees of freedom living on the $AdS_d$ slice, and they are coupled to the bulk fields in $AdS_{d+1}$. However, the $AdS_d$ brane is treated as a probe and does not back-reacts to the geometry. In \cite{Rastelli:2017ecj,Mazac:2018biw}, ``straight'' probe branes are considered in great detail where the $AdS_d$ slice is embedded in $AdS_{d+1}$ as the restriction to $z_d=0$. The setup corresponds to CFTs with a straight co-dimension 1 interface at $x_d=0$. One can then use method of images to take only half of the $AdS_{d+1}$ space with $z_d\geq 0$, and consider boundary CFTs defined on $x_d\geq 0$ with Dirichlet or Neumann boundary conditions, as was done in \cite{Mazac:2018biw}. Here we consider a slightly modified setup where the probe brane is a hemisphere in the Poincar\'e coordinates of $AdS_{d+1}$. It is related to the ``straight'' case by a conformal mapping. The method of images is similar in the spherical case (one can also first apply the method in the ``straight'' case and then perform the conformal mapping), and will not be elaborated here. We will therefore focus only on the probe brane case where the space continues beyond the interface.

As we pointed out in Section \ref{Sec:2.1}, the spherical boundary of a BCFT preserves the fixed embedding vector $N_B$ (\ref{NB}). In the bulk, the boundary of the BCFT is extended into the hemisphere 
\begin{equation}
N_B\cdot Z=\frac{1-z_0^2-\vec{z}^2}{2z_0}=0\;,
\end{equation}
which is the probe $AdS_d$ brane. This hemisphere coincides with the inversion hemisphere $\mathcal{H}$ defined in (\ref{defcalH}). Note that in the $\mathbb{RP}^d$ CFT case the fixed vector $N_c$ corresponds to a point in the bulk, while in the BCFT case the fixed vector $N_B$ is a normal vector defining a fixed co-dimension 1 surface. 

We can modify the Witten diagrams defined in Section \ref{Sec:3.2} to define their BCFT counterparts, by simply integrating over the whole hemisphere $\mathcal{H}$ instead of localizing on the north pole. For example, the BCFT one-point function is defined as 
\begin{equation}
\langle \mathcal{O}_\Delta(P)\rangle_B=\int_{\mathcal{H}}dZ_{\mathcal{H}} G^\Delta_{B\partial}(Z_{\mathcal{H}},P)\sim \frac{1}{|1-x^2|^\Delta}\;,
\end{equation}
which reproduces the correct structure (\ref{B1pt}). The two-point contact and exchange Witten diagrams are respectively defined as 
\begin{equation}\label{WBcon}
W^{con,0}_{B,\Delta_1,\Delta_2}(P_1,P_2)=\int_{\mathcal{H}}dZ_{\mathcal{H}} G^{\Delta_1}_{B\partial}(Z_{\mathcal{H}},P_1)G^{\Delta_2}_{B\partial}(Z_{\mathcal{H}},P_2)\;,
\end{equation}
\begin{equation}\label{WBexchange}
W^{exchange;bulk}_{B,\Delta}(P_1,P_2)=\int_{\mathcal{H}}dZ_{\mathcal{H}} \int_{AdS_{d+1}}dW G^{\Delta}_{BB}(Z_{\mathcal{H}},W) G^{\Delta_1}_{B\partial}(W,P_1)G^{\Delta_2}_{B\partial}(W,P_2)\;.
\end{equation}
It is easy to verify that a similar equation of motion identity relates the exchange Witten diagram to the contact Witten diagram
\begin{equation}
\mathbf{EOM}[W^{exchange;bulk}_{B,\Delta}]=W^{con,0}_{B,\Delta_1,\Delta_2}\;,
\end{equation}
by using similar arguments.
The integrals (\ref{WBcon}), (\ref{WBexchange}) can be mapped to the straight probe brane integrals studied  in \cite{Rastelli:2017ecj}, by parameterizing the hemisphere $\mathcal{H}$ with the following Poincar\'e coordinates\footnote{Here we switched to the $(-,+,-,+,\ldots,+)$ signature for $\mathbb{R}^{d,2}$. We also need to perform the Wick rotation $z_d\to iz_d$ to make the probe $AdS_d$ brane Euclidean in order to compare with \cite{Rastelli:2017ecj}.} 
\begin{equation}
Z_{\mathcal{H}}=\frac{1}{z_0}\left(z_d,0,\frac{1+(z_0^2-z_d^2+z_i^2)}{2},\frac{1-(z_0^2-z_d^2+z_i^2)}{2},z^i\right)\;,\quad i=1,\ldots,d-1\;.
\end{equation}
It then follows that these Witten diagrams are functions of the BCFT cross ratio $\xi$ defined in (\ref{defxi}), rather than the $\mathbb{RP}^d$ CFT cross ratio $\eta$ defined in (\ref{defeta}).

We can also define the geodesic Witten diagram similar to (\ref{defWgeo})
\begin{equation}
W^{geo;bulk}_{B,\Delta}(P_1,P_2)=\int_{\mathcal{H}}dZ_{\mathcal{H}}\int_{\gamma_{12}}d\gamma \; G^{\Delta}_{BB}(Z_{\mathcal{H}},\gamma)G_{B\partial}^{\Delta_1}(\gamma,P_1)G_{B\partial}^{\Delta_2}(\gamma,P_2)\;,
\end{equation}
as was first discussed in \cite{Rastelli:2017ecj}. A similar argument using the equation of motion operator shows that the geodesic Witten diagram is holographically dual to the bulk channel conformal block $g_{B,\Delta}^{bulk}(\xi)$ defined in (\ref{defgBbulk}).

Finally, in the probe brane setup we can define the boundary exchange Witten diagram 
\begin{equation}
W^{exchange;boundary}_{B,\Delta}(P_1,P_2)=\int_{\mathcal{H}}dZ_{1,\mathcal{H}}dZ_{2,\mathcal{H}}G^{\Delta,\mathcal{H}}_{BB}(Z_{1,\mathcal{H}},Z_{2,\mathcal{H}}) G^{\Delta_1}_{B\partial}(Z_{1,\mathcal{H}},P_1)G^{\Delta_2}_{B\partial}(Z_{2,\mathcal{H}},P_2)
\end{equation}
where an interface field localized on $\mathcal{H}$ with dimension $\Delta$ is exchanged via the $AdS_d$ propagator $G^{\Delta,\mathcal{H}}_{BB}(Z_{1,\mathcal{H}},Z_{2,\mathcal{H}})$. These diagrams have no analogue in the $\mathbb{RP}^d$ CFT case, since the inversion hemisphere is not a boundary and there is no extra degrees of freedom living on it.

\section{Two-point exchange Witten diagrams}\label{Sec:4}
In this section, we study in detail the properties of the two-point exchange Witten diagrams defined in the previous section. In Section \ref{Sec:evaluate2pt} we explicitly evaluate these diagrams. In Section \ref{Sec:cbdecomp} we study the conformal block decomposition of Witten diagrams in different channels. 

\subsection{Evaluating two-point exchange Witten diagrams}\label{Sec:evaluate2pt}
We will focus on the evaluation of the two-point exchange Witten diagram $W^{exchange}_\Delta$. The image diagram $\bar{W}^{exchange}_\Delta$ can be obtained from $W^{exchange}_\Delta$ via the crossing relation (\ref{calWWbcrossing}). The full $qAdS_{d+1}$ exchange Witten diagram $V^{exchange,\pm}_\Delta$ can then be assembled using (\ref{VasWWbar}). We first discuss the special case where $W^{exchange}_\Delta$ can be expressed as a finite sum of contact Witten diagrams. We then give the formula for the exchange diagram when the quantum numbers $\Delta_1$, $\Delta_2$, $\Delta$ are generic. Note that the calculation is exactly the same as doing only one of the two integrals in a scalar four-point exchange Witten diagram in $AdS_{d+1}$, since once we strip away the other two bulk-to-boundary propagators the integral becomes identical.

\vspace{0.5cm}
\noindent{\bf The truncated case} 
\vspace{0.1cm}

\noindent Let us first consider a special case when $\Delta_1+\Delta_2-\Delta\in 2\mathbb{Z}_+$. We can use the vertex identity for scalar exchange \cite{DHoker:1999mqo} (see also Appendix A of \cite{Goncalves:2019znr}) to write the integrated vertex as a finite sum of products of two bulk-to-boundary propagators with shifted conformal dimensions. The exchange diagram $W^{exchange}_\Delta$ then becomes 
\begin{equation}\label{Wtrunc}
\begin{split}
W^{exchange}_\Delta(P_1,P_2)={}&\sum_{k=k_{\min}}^{k_{\max}}a_k(\vec{x}_{12}^2)^{k-\Delta_2} G_{B\partial}^{k+\Delta_1-\Delta_2}(N_c,P_1)G_{B\partial}^k(N_c,P_2)\\
{}&=\sum_{k=k_{\min}}^{k_{\max}}a_k (\vec{x}_{12}^2)^{k-\Delta_2}  V^{con,0}_{k+\Delta_1-\Delta_2,k}(P_1,P_2)
\end{split}
\end{equation}
where 
\begin{equation}
\begin{split}
{}&k_{\min}=\frac{\Delta-\Delta_1+\Delta_2}{2}\;,\quad \quad k_{\max}=\Delta_2-1\;,\\
{}& a_{k-1}=\frac{(k-\frac{\Delta}{2}+\frac{\Delta_1-\Delta_2}{2})(k-\frac{d}{2}+\frac{\Delta}{2}+\frac{\Delta_1-\Delta_2}{2})}{(k-1)(k-1-\Delta_1+\Delta_2)}a_k\;,\\
{}& a_{\Delta_2-1}=\frac{1}{4(\Delta_1-1)(\Delta_2-1)}\;.
\end{split}
\end{equation}
Written in terms of the cross ratio, the exchange Witten diagram is a polynomial of $\eta^{-1}$
\begin{equation}
\mathcal{W}^{exchange}_\Delta(\eta)=\sum_{k=k_{\min}}^{k_{\max}}a_k\eta^{k-\Delta_2}\;.
\end{equation}

\vspace{0.5cm}
\noindent{\bf The general case} 
\vspace{0.1cm}

\noindent In the general case, we can still express the exchange Witten diagram as an infinite sum of contact Witten diagrams. The integral has already been computed in Appendix C of \cite{Zhou:2018sfz}. Here we review the derivation and the result in the language of $\mathbb{RP}^d$ CFT.

The main idea for evaluating the integral is to use the equation of motion relation (\ref{eomWtoV}) to write down a differential equation for the exchange Witten diagram. Written in terms of the cross ratio, the equation of motion identity becomes 
\begin{equation}
\mathbf{EOM}[\mathcal{W}^{exchange}_\Delta(\eta)]=1
\end{equation}
where the differential operator acts as 
\begin{equation}
\begin{split}
\mathbf{EOM}[\mathcal{G}(\eta)]={}&4\eta^2(\eta-1)\mathcal{G}''(\eta)+\eta(4(\eta-1)(\Delta_1+\Delta_2+1)+2d)\mathcal{G}'(\eta)\\
{}&+((\Delta-\Delta_1-\Delta_2)(-d+\Delta+\Delta_1+\Delta_2)+4\Delta_1\Delta_2\eta)\mathcal{G}(\eta)\;.
\end{split}
\end{equation}
The differential equation should be supplemented by two boundary conditions:
\begin{enumerate}
\item[1)] From the OPE limit $\eta\to0$, we know $\mathcal{W}^{exchange}_\Delta(\eta)$ should behave as $\eta^{\frac{\Delta-\Delta_1-\Delta_2}{2}}$.\footnote{Here we are assuming that $\Delta<\Delta_1+\Delta_2$ such that the single-trace contribution is leading.}
\item[2)] From the definition of the integral (\ref{defW}), $\mathcal{W}^{exchange}_\Delta(\eta)$ has to be smooth at $\eta=1$ (see \cite{DHoker:1998ecp}).
\end{enumerate}
The physical solution is a linear combination of the special solution
\begin{equation}
f(\eta)={}_3F_2\left(1,\Delta_1,\Delta_2;\frac{\Delta_1+\Delta_2-\Delta}{2}+1,\frac{\Delta_1+\Delta_2+\Delta-d}{2}+1;\eta\right)\;,
\end{equation}
and a homogeneous solution, the conformal block $g_\Delta(\eta)$, 
\begin{equation}\label{Wvalue}
\mathcal{W}^{exchange}_\Delta(\eta)=C_1 f(\eta)+C_2 g_\Delta(\eta)\;.
\end{equation}
The coefficients $C_1$, $C_2$ are given by 
\begin{equation}
\begin{split}
C_1={}&-\frac{1}{(\Delta_1+\Delta_2-\Delta)(\Delta_1+\Delta_2+\Delta-d)}\;,\\
C_2={}&\frac{\Gamma(\frac{\Delta+\Delta_1-\Delta_2}{2})\Gamma(\frac{\Delta-\Delta_1+\Delta_2}{2})\Gamma(\frac{-\Delta+\Delta_1+\Delta_2}{2})\Gamma(\frac{-d+\Delta+\Delta_1+\Delta_2}{2})}{4\Gamma(\Delta_1)\Gamma(\Delta_2)\Gamma(-\frac{d}{2}+\Delta+1)}\;.
\end{split}
\end{equation}
The ratio $\frac{C_1}{C_2}$ is precisely fixed by the condition that the solution is regular at $\eta=1$. We can also write (\ref{Wvalue}) as two infinite series in $\eta$. Using the definition (\ref{defV0a}), (\ref{defV0b}) for the contact Witten diagram, we can then write $W^{exchange}_\Delta$ as two infinite sums of contact Witten diagrams 
\begin{equation}
W^{exchange}_\Delta=\sum_{i=0}^\infty (\vec{x}_{12}^2)^i P_i V^{con,0}_{\Delta_1+i,\Delta_2+i}+\sum_{i=0}^\infty (\vec{x}_{12}^2)^{\frac{\Delta-\Delta_1-\Delta_2+2i}{2}}Q_i V^{con,0}_{\frac{\Delta+\Delta_1-\Delta_2}{2}+i, \frac{\Delta-\Delta_1+\Delta_2}{2}+i}
\end{equation}
where the coefficients $P_i$ and $Q_i$ are given by
\begin{equation}
P_i=\frac{(\Delta_1)_i (\Delta_2)_i}{(\Delta -\Delta_1-\Delta_2) (-d+\Delta +\Delta_1+\Delta_2) \left(\frac{-\Delta +\Delta_1+\Delta_2+2}{2}\right)_i \left(\frac{-d+\Delta +\Delta_1+\Delta_2+2}{2}\right)_i}\;,
\end{equation}
and
\begin{equation}
\begin{split}
Q_i={}&\frac{(-1)^i \Gamma \left(\frac{d-2 i-2\Delta}{2}\right)\sin \left(\frac{\pi  (d-2 \Delta )}{2} \right)\Gamma \left(\frac{-d+\Delta +\Delta_1+\Delta_2}{2}\right)}{4 \pi  \Gamma (i+1)\Gamma (\Delta_1) \Gamma (\Delta_2)}\\
{}&\times \frac{\Gamma \left(\frac{\Delta -\Delta_1+\Delta_2}{2}\right) \Gamma \left(\frac{\Delta +\Delta_1-\Delta_2}{2}\right) \Gamma \left(\frac{-\Delta +\Delta_1+\Delta_2}{2}\right)\Gamma \left(\frac{-\Delta +\Delta_1-\Delta_2+2}{2}\right)  \Gamma \left(\frac{-\Delta -\Delta_1+\Delta_2+2}{2}\right) }{\Gamma \left(\frac{-\Delta +\Delta_1-\Delta_2-2 i+2}{2}\right)\Gamma \left(\frac{-\Delta -\Delta_1+\Delta_2-2 i+2}{2}\right)}\;.
\end{split}
\end{equation}
When $\Delta_1+\Delta_2-\Delta=2\mathbb{Z}_+$ the infinite sum truncates and reduces to (\ref{Wtrunc}).

Finally, let us examine the behavior of the exchange Witten diagrams at $\eta\to\infty$. By using the equation of motion identities (\ref{eomWtoV}) and the behavior of the contact Witten diagram (\ref{VconRegge}), we find 
\begin{equation}\label{WWbarRegge}
\mathcal{W}^{exchange}_\Delta(\eta)\to \eta^{-1}\;, \bar{\mathcal{W}}^{exchange}_\Delta(\eta)\to \eta^{-1}\;,\quad \text{for  } \eta\to\infty\;. 
\end{equation}

\subsection{Conformal block decomposition of Witten diagrams}\label{Sec:cbdecomp}
\vspace{0.5cm}
\noindent{\bf The contact Witten diagram} 
\vspace{0.1cm}

\noindent We now study the conformal block decomposition of two-point Witten diagrams. We start with a warmup case, namely the zero-derivative contact Witten diagram (\ref{defV0b}). It is straightforward to show that $\mathcal{V}^{con,0}_{\Delta_1,\Delta_2}(\eta)$ can be written as an infinite sum over double-trace conformal blocks
\begin{equation}\label{cfdecomVcon0}
\mathcal{V}^{con,0}_{\Delta_1,\Delta_2}(\eta)=\sum_{n}a_n g_{\Delta_n^{d.t.}}(\eta)=\sum_{n}a_n \bar{g}_{\Delta_n^{d.t.}}(\eta)
\end{equation} 
where $\Delta_n^{d.t.}\equiv \Delta_1+\Delta_2+2n$, and 
\begin{equation}\label{an}
a_n=\frac{(-1)^n \Gamma (n+\Delta_1) \Gamma (n+\Delta_2) \Gamma \left(-\frac{d}{2}+n+\Delta_1+\Delta_2\right)}{\Gamma (\Delta_1) \Gamma (\Delta_2) \Gamma (n+1) \Gamma \left(-\frac{d}{2}+\Delta_n^{d.t.}\right)}\;.
\end{equation}
The exchanged operators in each channel correspond to double-trace operators of the schematic form $:\mathcal{O}_1\square^n\mathcal{O}_2:$ with $n=0,1,\ldots$.

\vspace{0.5cm}
\noindent{\bf Exchange Witten diagram in the direct channel} 
\vspace{0.1cm}

\noindent Let us now consider the conformal block decomposition of the exchange Witten diagram (\ref{defW}) in the same channel. The Witten diagram can be written as a sum of a single-trace conformal block with dimension $\Delta$, dual to the exchanged field, and infinitely many double-trace conformal blocks  
\begin{equation}\label{Wdecomdir}
\mathcal{W}^{exchange}_\Delta(\eta)=A\, g_\Delta(\eta)+\sum_{n}A_n g_{\Delta_n^{d.t.}}(\eta)\;.
\end{equation}
The single-trace OPE coefficient can be extracted from the small $\eta$ expansion of (\ref{Wvalue}), and is associated to the term with the behavior $\eta^{\frac{\Delta-\Delta_1-\Delta_2}{2}}$
\begin{equation}\label{WAcoe}
A=\frac{\Gamma \left(\frac{\Delta +\Delta_1-\Delta_2}{2}\right) \Gamma \left(\frac{\Delta -\Delta_1+\Delta_2}{2}\right) \Gamma \left(\frac{-\Delta +\Delta_1+\Delta_2}{2}\right) \Gamma \left(\frac{-d+\Delta +\Delta_1+\Delta_2}{2}\right)}{4 \Gamma (\Delta_1) \Gamma (\Delta_2) \Gamma \left(-\frac{d}{2}+\Delta +1\right)}\;.
\end{equation}
To extract the double-trace OPE coefficients, we use the equation of motion identity (\ref{eomWtoV}). Note that the $\mathbf{EOM}$ operator annihilates the single-trace conformal block $g_\Delta(\eta)$, while multiplies the double-trace conformal blocks with constants
\begin{equation}
\begin{split}
\mathbf{EOM}[g_\Delta(\eta)]={}&0\;,\\
\mathbf{EOM}[g_{\Delta_n^{d.t.}}(\eta)]={}&(\Delta(\Delta-d)-\Delta_n^{d.t.}(\Delta_n^{d.t.}-d))g_{\Delta_n^{d.t.}}(\eta)\;.
\end{split}
\end{equation}
Using the conformal block decomposition (\ref{cfdecomVcon0}) of the contact Witten diagram, we find 
\begin{equation} \label{WAnDDT}
A_n=\frac{a_n}{\Delta(\Delta-d)-\Delta_n^{d.t.}(\Delta_n^{d.t.}-d)}\;.
\end{equation}
Using these OPE coefficients, we can further expand the conformal blocks to obtain a small $\eta$ expansion for the exchange Witten diagram. This expansion can be compared with the expansion of (\ref{Wvalue}), which provides a consistency check of our results. By crossing symmetry, we also have 
\begin{equation}
\bar{\mathcal{W}}^{exchange}_\Delta(\eta)=A\, \bar{g}_\Delta(\eta)+\sum_{n}A_n \bar{g}_{\Delta_n^{d.t.}}(\eta)\;.
\end{equation}

\vspace{0.5cm}
\noindent{\bf Exchange Witten diagram in the crossed channel} 
\vspace{0.1cm}

\noindent Finally we consider the conformal block decomposition of the exchange Witten diagram in the crossed channel. The decomposition consists of crossed channel double-trace conformal blocks only
\begin{equation}
\mathcal{W}^{exchange}_\Delta(\eta)=\sum_n B_n \bar{g}_{\Delta_n^{d.t.}}(\eta)\;.
\end{equation}
To work out the decomposition coefficients, we will generalize the recursive techniques developed in \cite{Zhou:2018sfz}. We apply the equation of motion relation (\ref{eomWtoV}) to turn the exchange Witten diagram into a contact Witten diagram, which has already been decomposed into the crossed channel in (\ref{cfdecomVcon0}). On the other hand, the action of the $\mathbf{EOM}$ operator on $\bar{g}_{\Delta_n^{d.t.}}$ admits a simple three-term recursion relation
\begin{equation}\label{cfblockrecur}
\mathbf{EOM}[\bar{g}_{\Delta_n^{d.t.}}]=\mu_n \bar{g}_{\Delta_{n-1}^{d.t.}}+\nu_n \bar{g}_{\Delta_n^{d.t.}}+\rho_n \bar{g}_{\Delta_{n+1}^{d.t.}}
\end{equation}
where 
\begin{eqnarray}
\nonumber \mu_n=&&(\Delta_1+\Delta_2-\Delta_n^{d.t.}) (d-\Delta_1-\Delta_2+\Delta_n^{d.t.}-2)\;,\\
\nonumber \nu_n=&&\frac{(d-2 \Delta_1-2 \Delta_2+2) (3 d-2 \Delta_1-2 \Delta_2-2) (d+2 \Delta_1-2 \Delta_2-2) (d-2 \Delta_1+2 \Delta_2-2)}{8 (d-2 \Delta_n^{d.t.}-2) (d-2 \Delta_n^{d.t.}+2)}\\
\nonumber &&-\left(\Delta_1-\frac{d}{2}\right)^2-\left(\Delta_2-\frac{d}{2}\right)^2+\frac{1}{2} \left(\Delta_n^{d.t.}-\frac{d}{2}\right)^2+d-\frac{1}{2}+\Delta  (\Delta -d)\;,\\
\rho_n=&&\frac{\left((\Delta_1-\Delta_2)^2-{\Delta_n^{d.t.}}^2\right) (d-\Delta_1-\Delta_2-\Delta_n^{d.t.}) (2 d-\Delta_1-\Delta_2-\Delta_{n+1}^{d.t.})}{(d-2 (\Delta_n^{d.t.}+1))^2}\\
\nonumber &&\times\frac{(d+\Delta_1-\Delta_2-\Delta_{n+1}^{d.t.}) (d-\Delta_1+\Delta_2-\Delta_{n+1}^{d.t.})}{(d-2 \Delta_n^{d.t.})(d-2\Delta_{n+1}^{d.t.})}\;.
\end{eqnarray}
Note that for $n=0$, the coefficient $\mu_0$ vanishes. Therefore the action of the $\mathbf{EOM}$ operator preserves the double-trace spectrum. Using (\ref{cfblockrecur}), (\ref{Wvalue}) and (\ref{cfdecomVcon0}), we obtain the following inhomogeneous recursion relation for the decomposition coefficients 
\begin{equation} \label{BnRecursion}
\rho_{n-1}B_{n-1}+\nu_n B_n+\mu_{n+1} B_{n+1}=a_n\;.
\end{equation}
For $n = 0$, the equation remains the same just without the $\rho_n$ piece. The coefficients $B_n$ with $n>0$ can be recursively solved, after specifying the boundary condition $B_0$, which is extracted from the $\eta\to1$ limit of the exchange diagram (\ref{Wvalue})
\begin{equation}\label{B0BoundaryCondition}
\begin{split}
&B_0=\mathcal{W}^{exchange}_\Delta( \eta \rightarrow 1) \\
&= \frac{\Gamma\left(\frac{\Delta_1 + \Delta_2 - \Delta}{2} \right) \Gamma\left(\frac{\Delta_1 + \Delta_2 - d + \Delta}{2} \right) \Gamma\left( 1 - \frac{d}{2} \right) }{4 \Gamma (\Delta_{2})}  \Bigg[\frac{\Gamma \left(\frac{\Delta + \Delta_1 - \Delta_2}{2}\right)\Gamma \left(\frac{\Delta - \Delta_1 + \Delta_2}{2}\right)}{\Gamma\left(\Delta_1 \right) \Gamma \left(\frac{ 2 - d + \Delta + \Delta_1 - \Delta_2}{2}\right) \Gamma \left(\frac{ 2 - d + \Delta - \Delta_1 + \Delta_2}{2}\right) } \\
& -  \ {}_3 \tilde{F}_2 \left( \frac{2 + \Delta_1 - \Delta_2 - \Delta}{2}, \frac{2 + \Delta_1 - \Delta_2 - d + \Delta}{2}, 1 - \frac{d}{2}; 2 - \frac{d}{2}, 1 - \frac{d}{2} + \Delta_{1}; 1 \right)   \Bigg]
\end{split}
\end{equation}
where ${}_3 \tilde{F}_2$ is the regularized hypergeometric function defined by 
\begin{equation}
{}_3 \tilde{F}_2 (a_1, a_2, a_3; b_1, b_2; z) = \frac{{}_3 F_2 (a_1, a_2, a_3; b_1, b_2; z)}{\Gamma (b_1) \Gamma (b_2)}. 
\end{equation}
By crossing symmetry, the conformal block decomposition for $\mathcal{W}^{exchange}_\Delta$ also gives 
\begin{equation}
\bar{\mathcal{W}}^{exchange}_\Delta(\eta)=\sum_n B_n g_{\Delta_n^{d.t.}}(\eta)\;.
\end{equation}

\section{Dimensional reduction}\label{Sec:5}
In \cite{Zhou:2020ptb} it was pointed out that a large class of Witten diagram recursion relations can be obtained from conformal block recursion relations, essentially by just replacing conformal blocks with the corresponding exchange Witten diagrams. This idea was demonstrated for four-point functions in generic CFTs, and two-point functions in CFTs with boundaries. Here we generalize similar statements to two-point functions in CFTs on $\mathbb{RP}^d$, and we will focus on the dimensional reduction relations.
\subsection{Reduction for conformal blocks}
Let us recall the relation between $\mathbb{RP}^d$ CFT conformal blocks and BCFT conformal blocks in the bulk channel (see (\ref{defg}) and (\ref{defgBbulk}))
\begin{equation}
g_\Delta(\eta)=(-1)^{\frac{\Delta-\Delta_1-\Delta_2}{2}}g_{B,\Delta}^{bulk}(-\eta)\;.
\end{equation}
The dimensional reduction formulae derived in \cite{Zhou:2020ptb} for $g_{B,\Delta}^{bulk}(\xi)$ therefore can be straightforwardly transformed into those for $g_\Delta(\eta)$. We have the following relation between conformal blocks in $d$ and $d-1$ dimensions
\begin{equation}\label{gdtodm1}
g^{(d)}_\Delta(\eta)=\sum_{j=0}^\infty \alpha^{(d)}_j(\Delta) g^{(d-1)}_{\Delta+2j}(\eta)
\end{equation}
where we have used the superscript to emphasize the dimensional dependence, and 
\begin{equation}
\alpha^{(d)}_j(\Delta)=\frac{\Gamma \left(j+\frac{1}{2}\right) \left(\frac{1}{2} (\Delta +\Delta_1-\Delta_2)\right)_j \left(\frac{1}{2} (\Delta -\Delta_1+\Delta_2)\right)_j}{\sqrt{\pi } j! \left(\frac{1}{2} (-d+2 \Delta +2)\right)_j \left(j+\frac{1}{2} (-d+2 \Delta -1)+1\right)_j}\;.
\end{equation}
Moreover, a conformal block in $d-2$ dimensions can be expressed in terms of only two conformal blocks in $d$ dimensions
\begin{equation}\label{gdtodm2}
g^{(d-2)}_\Delta(\eta)=g^{(d)}_\Delta(\eta)+\beta(\Delta) g^{(d)}_{\Delta+2}(\eta)
\end{equation}
where
\begin{equation}
\beta(\Delta)=-\frac{(\Delta +\Delta_1-\Delta_2) (\Delta -\Delta_1+\Delta_2)}{(d-2 \Delta -4) (d-2 \Delta -2)}\;.
\end{equation}
Using $\bar{g}^{(d)}_\Delta(\eta)=g^{(d)}_\Delta(1-\eta)$, we also obtain similar dimensional reduction formulae for the image channel conformal blocks $\bar{g}^{(d)}_\Delta(\eta)$. 

Let us comment that the recursion relation (\ref{gdtodm2}) is quite special, as it involves only finitely many terms. In fact, the inverse relation which expresses a $d$ dimensional conformal block in terms of $d-2$ dimensional blocks, contains infinitely many conformal blocks. A similar relation of (\ref{gdtodm2}) was first derived in \cite{Kaviraj:2019tbg} for conformal blocks for four-point functions in CFTs without boundaries. The identity expresses a $d-2$ dimensional conformal block in terms of the linear combination of five conformal blocks in $d$ dimensions. The existence of such a recursion was explained in terms of a $OSp(d+1,1|2)$ Parisi-Sourlas supersymmetry \cite{Parisi:1979ka} in a $d$ dimensional SCFT, which upon dimensional reduction gives rise to a non-supersymmetric CFT in $d-2$ dimensions. The relation (\ref{gdtodm2}) we wrote down here parallels the five-term relation in \cite{Kaviraj:2019tbg}, and therefore suggests that a similar story of Parisi-Sourlas supersymmetry and dimensional reduction can also be extended to CFTs on real projective space.

\subsection{Reduction for exchange Witten diagrams}
Similar to the observations in \cite{Zhou:2020ptb}, the recursion relations (\ref{gdtodm1}) and (\ref{gdtodm2}) can also be extended to imply relations for exchange Witten diagrams.  Let us rescale the exchange Witten diagrams such that the single-trace conformal blocks appear with unit coefficients
\begin{equation}\label{defPolyakovb}
\mathcal{P}_\Delta(\eta)=\frac{1}{A}\mathcal{W}^{exchange}_\Delta(\eta)\;,\quad \bar{\mathcal{P}}_\Delta(\eta)=\frac{1}{A}\bar{\mathcal{W}}^{exchange}_\Delta(\eta)
\end{equation}
where $A$ is the coefficient of the single-trace conformal block (\ref{WAcoe}). We claim that we have the following dimensional reduction formulae
\begin{equation}\label{Pdtodm1}
\mathcal{P}^{(d)}_\Delta(\eta)=\sum_{j=0}^\infty \alpha^{(d)}_j(\Delta) \mathcal{P}^{(d-1)}_{\Delta+2j}(\eta)\;,
\end{equation}
\begin{equation}\label{Pdtodm2}
\mathcal{P}^{(d-2)}_\Delta(\eta)=\mathcal{P}^{(d)}_\Delta(\eta)+\beta(\Delta)\mathcal{P}^{(d)}_{\Delta+2}(\eta)\;.
\end{equation}
Similar relations also hold for the mirror channel exchange Witten diagram upon replacing $\mathcal{P}^{(d)}_\Delta(\eta)$ with $\bar{\mathcal{P}}^{(d)}_\Delta(\eta)$. In \cite{Zhou:2020ptb}, similar Witten diagram identities were proven by using simple Mellin space arguments. Unfortunately, the same arguments cannot be used here. We note that although a Mellin representation formalism  can be developed for $\mathbb{RP}^d$ correlators similar to the BCFT case \cite{Rastelli:2017ecj}, it is not suitable for holographic correlators. To see this, we recall that contact Witten diagrams are polynomials of the cross ratio, and their Mellin transform are ill-defined. Nevertheless, we can still prove (\ref{Pdtodm1}) and  (\ref{Pdtodm2}) in position space by using the conformal block decomposition in the direct channel. 

Let us denote the decomposition of the exchange Witten diagrams as
\begin{equation}
\mathcal{P}^{(d)}_\Delta(\eta)=g^{(d)}_\Delta(\eta)+\sum_{n=0}^\infty \mu^{(d)}_n(\Delta)g^{(d)}_{\Delta_{n}^{d.t.}}(\eta)
\end{equation}
where $\mu^{(d)}_n(\Delta)=\frac{A_n}{A}$ in relation to (\ref{Wdecomdir}). Substituting this decomposition into (\ref{Pdtodm2}), we find that the single-trace conformal blocks on both sides cancel thanks to (\ref{gdtodm2}). The double-trace conformal blocks must also match, provided 
\begin{equation}
\sum_{n=0}^\infty \mu^{(d-2)}_n(\Delta)g^{(d-2)}_{\Delta_{n}^{d.t.}}(\eta)=\sum_{n=0}^\infty \mu^{(d)}_n(\Delta)g^{(d)}_{\Delta_{n}^{d.t.}}(\eta)+\beta(\Delta) \sum_{n=0}^\infty \mu^{(d)}_n(\Delta+2)g^{(d)}_{\Delta_{n}^{d.t.}}(\eta)\;.
\end{equation}
We can use (\ref{gdtodm2}) again to turn $g^{(d-2)}_{\Delta_{n}^{d.t.}}(\eta)$ into $g^{(d)}_{\Delta_{n}^{d.t.}}(\eta)$, and arrive at the following condition
\begin{equation}
\mu^{(d-2)}_n(\Delta)+\beta(\Delta_{n-1})\mu^{(d-2)}_{n-1}(\Delta)=\mu^{(d)}_n(\Delta)+\beta(\Delta)\mu^{(d)}_n(\Delta+2)\;.
\end{equation}
This identity can be straightforwardly verified, using the explicit expressions for $\mu^{(d)}_n(\Delta)$ and $\beta(\Delta)$. 

We can proceed similarly for the relation (\ref{Pdtodm1}). The single-trace operators again drop out because of the conformal block recursion relation. The double-trace coefficients need to be constrained, and the condition reads
\begin{equation}
\sum_{m+j=n} \mu^{(d)}_m(\Delta) \alpha^{(d)}_j(\Delta_m)=\sum_{k=0}^\infty \alpha^{(d)}_k(\Delta)\mu^{(d-1)}_n(\Delta+2k)\;.
\end{equation}
The infinite sum on the r.h.s. makes it difficult to check analytically, we can nevertheless numerically check this identity. 

Another nontrivial crosscheck is to use (\ref{Pdtodm1}) twice to reproduce (\ref{Pdtodm2}). It is not difficult to find 
\begin{equation}
\begin{split}
\mathcal{P}^{(d)}_\Delta(\eta)={}&\sum_{j,k}\alpha^{(d)}_j(\Delta)\alpha^{(d-1)}_k(\Delta+2j)\mathcal{P}^{(d-2)}_{\Delta+2j+2k}(\eta)\\
={}&\sum_{n=0}^\infty \frac{\left(\frac{\Delta +\Delta_1-\Delta_2}{2}\right)_n \left(\frac{\Delta -\Delta_1+\Delta_2}{2}\right)_n}{\left(-\frac{d}{2}+\Delta +1\right)_{2 n}}\mathcal{P}^{(d-2)}_{\Delta+2n}(\eta)\;.
\end{split}
\end{equation}
Using this identity in (\ref{Pdtodm2}), one can straightforwardly verify that the relation is valid.

\section{An analytic bootstrap approach for $\mathbb{RP}^d$ CFTs}\label{Sec:6}
In this section, we present an analytic bootstrap approach for studying CFTs on $\mathbb{RP}^d$. Part of our discussions forms a close analogy of the analysis for BCFT two-point functions in \cite{Mazac:2018biw} (see also the related works \cite{Mazac:2018ycv,Kaviraj:2018tfd,Mazac:2019shk,Caron-Huot:2020adz}). In Section \ref{Sec:basisfunctional} we argue that the double-trace conformal blocks in both the bulk channel and the mirror channels form a complete basis for two-point correlators. Their duals give a basis for analytic functionals, and we will explicitly construct their actions using exchange Witten diagrams. In Section \ref{Sec:dispersion} we give another construction of the functionals from the dispersion relation. We apply these analytic functionals in Section \ref{Sec:testfunctionals}, where we obtain the $\epsilon$-expansion result of $\mathbb{RP}^d$  $\varphi^4$ theory to $\epsilon^2$ order. We also perform an independent field theory check of our results in Section \ref{Sec:CFTEoM}.

\subsection{Space of functions, double-trace basis, and dual basis}\label{Sec:basisfunctional}
The study of Witten diagrams in Section \ref{Sec:4} motivates us to propose a natural basis for two-point functions. As we have seen, the exchange Witten diagrams admit the following decompositions in two channels 
\begin{equation}\label{Wtwochannels}
\mathcal{W}^{exchange}_\Delta(\eta)=A\, g_\Delta(\eta)+\sum_{n}A_n g_{\Delta_n^{d.t.}}(\eta)=\sum_n B_n \bar{g}_{\Delta_n^{d.t.}}(\eta)\;,
\end{equation} 
\begin{equation}\label{Wbartwochannels}
\bar{\mathcal{W}}^{exchange}_\Delta(\eta)=A\, \bar{g}_\Delta(\eta)+\sum_{n}A_n \bar{g}_{\Delta_n^{d.t.}}(\eta)=\sum_n B_n g_{\Delta_n^{d.t.}}(\eta)\;.
\end{equation} 
These identities show that any conformal blocks $g_\Delta(\eta)$, $\bar{g}_\Delta(\eta)$ with generic conformal dimension $\Delta$ can be expressed as linear combinations of double-trace conformal blocks $g_{\Delta_n^{d.t.}}(\eta)$, $\bar{g}_{\Delta_n^{d.t.}}(\eta)$ in both channels. This fact, loosly speaking, implies that $\{g_{\Delta_n^{d.t.}}(\eta),\bar{g}_{\Delta_n^{d.t.}}(\eta)\}$ form a new basis. 

To phrase our statement more precisely, we need to define the space of functions $\mathcal{U}$ for the correlators $\mathcal{G}(\eta)$ which we are considering. We define $\mathcal{U}$ to be the space with the following ``Regge'' behavior
\begin{equation}\label{spaceU}
\mathcal{G}(\eta)\in \mathcal{U}\;,\;\text{if  }\;\; |\mathcal{G}|\lesssim |\eta|^{-\epsilon}\;,\;\text{when  }\;\; \eta\to\infty
\end{equation}
where $\epsilon$ is an infinitesimal positive number. For example, the mean field theory two-point function 
\begin{equation}\label{Reggebehavior}
\langle\phi_\pm(x_1)\phi_\pm(x_2)\rangle=\frac{1}{(1+x_1^2)^{\Delta_\phi}(1+x_2^2)^{\Delta_\phi}}(\eta^{-\Delta_\phi}\pm(1-\eta)^{-\Delta_\phi})\;,
\end{equation}
belongs to this space when $\Delta_\phi>0$\;. The conformal blocks $g_\Delta(\eta)$, $\bar{g}_\Delta(\eta)$ are also in this space if the external dimensions $\min\{\Delta_1,\Delta_2\}>0$. On the other hand, the contact Witten diagrams are not in this space (see (\ref{VconRegge})). This avoids having relations among the basis vectors $\{g_{\Delta_n^{d.t.}}(\eta),\bar{g}_{\Delta_n^{d.t.}}(\eta)\}$, as a contact Witten diagram can be decomposed into only double-trace conformal blocks in either channel. Note that in the BCFT case, it was proven that two-point correlators in any unitary theory have a bounded Regge behavior when $\xi\to -1$ \cite{Mazac:2018biw}. The proof exploits the positivity of the decomposition coefficients in the boundary channel. By contrast, in the case at hand here of $\mathbb{RP}^d$ CFTs, positivity is not {\it a priori} guaranteed in either channel even when the theory is unitary. The Regge behavior requirement (\ref{spaceU}) is therefore imposed by hand. 

We claim that the double-trace conformal blocks  $\{g_{\Delta_n^{d.t.}},\bar{g}_{\Delta_n^{d.t.}}\}$ form a basis for the space $\mathcal{U}$. A basis for the dual space $\mathcal{U}^*$ is given by the set of functionals $\{\omega_m,\bar{\omega}_m\}$, defined by dualizing the double-trace basis
\begin{equation}\label{orthonormal}
\begin{split}
{}&\omega_m(g_{\Delta_n^{d.t.}})=\delta_{mn}\;,\quad \omega_m(\bar{g}_{\Delta_n^{d.t.}})=0\;,\\
{}&\bar{\omega}_m(g_{\Delta_n^{d.t.}})=0\;,\quad \bar{\omega}_m(\bar{g}_{\Delta_n^{d.t.}})=\delta_{mn}\;.
\end{split}
\end{equation}
Although we do not have a general proof for this proposal (except for the $d=2$ case where we prove in Section \ref{Sec:dispersion} from the dispersion relation), we will provide ample evidence which supports this conjecture.

The action of the basis functionals can be read off from the conformal block decompositions of exchange Witten diagrams. Acting on (\ref{Wtwochannels}) with $\omega_m$ and use the orthonormal relation (\ref{orthonormal}), we get 
\begin{equation} \label{FunctionalActionBlockDirect}
\omega_m(g_\Delta)=-\frac{A_m}{A}\;.
\end{equation}
Acting with $\bar{\omega}_m$, we find
\begin{equation}
\bar{\omega}_m(g_\Delta)=\frac{B_m}{A}\;.
\end{equation}
The action of the functionals on the image channel conformal block $\bar{g}_\Delta(\eta)$ can be obtained from (\ref{Wbartwochannels}), and is related to the action on $g_\Delta(\eta)$ by crossing symmetry
\begin{equation}
\omega_m(\bar{g}_\Delta)=\bar{\omega}_m(g_\Delta)\;,\quad \bar{\omega}_m(\bar{g}_\Delta)=\omega_m(g_\Delta)\;.
\end{equation}
Let us consider the action of the functionals on a two-point function $\mathcal{G}\in \mathcal{U}$ with the following conformal block decomposition
\begin{equation}\label{calGexpandinggbar}
\mathcal{G}(\eta)=\sum_{k} \mu_{12k} g_{\Delta_{\mathcal{O}_k}}(\eta)=\pm \sum_{k} \mu_{12k} \bar{g}_{\Delta_{\mathcal{O}_k}}(\eta)\;.
\end{equation}
Applying the basis functionals allow us to extract the complete set of constraints in terms of the sum rules 
\begin{equation}\label{sumrulea}
\sum_{k} \mu_{12k} \omega_n(g_{\Delta_{\mathcal{O}_k}})\mp \sum_{k} \mu_{12k} \omega_n(\bar{g}_{\Delta_{\mathcal{O}_k}})=0\;,
\end{equation}
\begin{equation}\label{sumruleb}
\sum_{k} \mu_{12k} \bar{\omega}_n(g_{\Delta_{\mathcal{O}_k}})\mp \sum_{k} \mu_{12k} \bar{\omega}_n(\bar{g}_{\Delta_{\mathcal{O}_k}})=0\;.
\end{equation}

The exchange Witten diagrams can also be viewed as the Polyakov-Regge blocks in the Polyakov-style bootstrap \cite{Polyakov:1974gs,Gopakumar:2016wkt,Gopakumar:2016cpb,Dey:2016mcs,Dey:2017fab,Gopakumar:2018xqi,Mazac:2018ycv,Kaviraj:2018tfd,Mazac:2018biw,Mazac:2019shk,Ferrero:2019luz,Penedones:2019tng,Sleight:2019ive,Caron-Huot:2020adz}, and can be used  as a new decomposition basis. In terms of the rescaled exchanged Witten diagrams (\ref{defPolyakovb}), two-point function in (\ref{calGexpandinggbar}) can be rewritten as 
\begin{equation}\label{Pblockexpand}
\mathcal{G}(\eta)=\sum_k \mu_{12k} (\mathcal{P}_{\Delta_{\mathcal{O}_k}}(\eta)\pm \bar{\mathcal{P}}_{\Delta_{\mathcal{O}_k}}(\eta))\;.
\end{equation}
To prove this relation, we can express (\ref{Wtwochannels}), (\ref{Wbartwochannels}) in terms of functional actions, and substitute in (\ref{Pblockexpand}) with 
\begin{equation}\label{Ping}
\mathcal{P}_\Delta(\eta)=g_\Delta(\eta)-\sum_n \omega_n(g_\Delta) g_{\Delta_n^{d.t.}}(\eta)\;,
\end{equation}
\begin{equation}\label{Pbaring}
\bar{\mathcal{P}}_\Delta(\eta)=\sum_n \omega_n(\bar{g}_\Delta) g_{\Delta_n^{d.t.}}(\eta)\;.
\end{equation}
We now have 
\begin{equation}
\mathcal{G}(\eta)=\sum_{k} \mu_{12k} g_{\Delta_{\mathcal{O}_k}}(\eta)-\sum_{k} \mu_{12k}\bigg(\sum_n \omega_n(g_{\Delta_{\mathcal{O}_k}}) g_{\Delta_n^{d.t.}}(\eta)\mp\sum_n \omega_n(\bar{g}_{\Delta_{\mathcal{O}_k}}) g_{\Delta_n^{d.t.}}(\eta)\bigg)\;.
\end{equation}
Interchanging the action of the functionals and the sums, the second term becomes  
\begin{equation}
\sum_n \omega_n\bigg(\sum_{k} \mu_{12k}(g_{\Delta_{\mathcal{O}_k}}(\eta)\mp\bar{g}_{\Delta_{\mathcal{O}_k}}(\eta))\bigg) g_{\Delta_n^{d.t.}}(\eta)\;,
\end{equation}
and vanishes because of the crossing equation (\ref{calGexpandinggbar}). We have therefore proven the equivalence between (\ref{Pblockexpand}) and the first identity in (\ref{calGexpandinggbar}). To prove the second identity, we just need to decompose the Polyakov-Regge blocks in the image channel
\begin{equation}\label{Pingbar}
\mathcal{P}_\Delta(\eta)=\sum_n \bar{\omega}_n(g_\Delta) \bar{g}_{\Delta_n^{d.t.}}(\eta)\;,
\end{equation}
\begin{equation}\label{Pbaringbar}
\bar{\mathcal{P}}_\Delta(\eta)=\bar{g}_\Delta(\eta)-\sum_n \bar{\omega}_n(\bar{g}_\Delta) \bar{g}_{\Delta_n^{d.t.}}(\eta)\;.
\end{equation}

The alternative decomposition (\ref{Pblockexpand}) can also be taken as the starting point for analytic bootstrap. In a generic interacting CFT, we do not expect operators with precise double-trace dimensions. However, if we use the the conformal block decompositions (\ref{Ping}), (\ref{Pbaring}), (\ref{Pingbar}), (\ref{Pbaringbar}) of the Polyakov-Regge blocks, we would encounter spurious double-trace operators in (\ref{Pblockexpand}). The requirement that these spurious operators should cancel gives sum   rules for the OPE coefficients, which are identical to the conditions (\ref{sumrulea}), (\ref{sumruleb}).

\subsection{Functionals from dispersion relation}\label{Sec:dispersion}
As was pointed in \cite{Mazac:2019shk}, analytic functionals and the dispersion relation for correlators \cite{Carmi:2019cub} (see also \cite{Bissi:2019kkx}) are closely related. In particular, the kernel of the dispersion relation can be viewed as the generating function for the kernels of the analytic functional in their integral representation. Here we will demonstrate that a similar relation holds for $\mathbb{RP}^d$ CFTs by re-deriving the basis identified in Section \ref{Sec:basisfunctional} and constructing the dual functionals. For  simplicity, we will only focus on $d=2$ and set $\Delta_1=\Delta_2=\Delta_\phi$.

\begin{figure}[htbp]
\begin{center}
\includegraphics[width=0.65\textwidth]{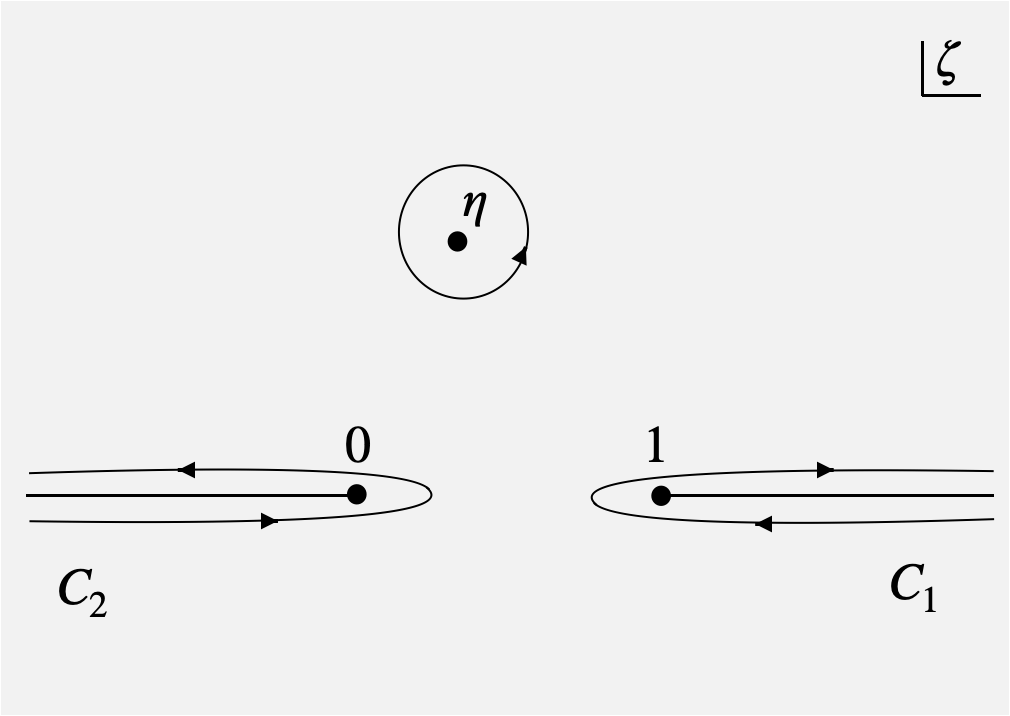}
\caption{An illustration of the integral contours in the dispersion relation.}
\label{Fig:contours}
\end{center}
\end{figure}

We start with the Cauchy's integral formula 
\begin{equation}
\mathcal{G}(\eta)=\oint \frac{d \zeta}{2\pi i} \frac{\mathcal{G}(\zeta)}{\zeta-\eta}\;,
\end{equation}
with a contour encircling the point $\zeta=\eta$. We can deform the contour, and wrap it around the two branch cuts $[1,\infty)$, $(-\infty,0]$. We will denote these two deformed contours respectively as $C_1$ and $C_2$, as is illustrated in Figure \ref{Fig:contours}. Since we have assumed that the correlator has the Regge behavior (\ref{spaceU}), we can safely drop the arc at infinity. We therefore obtain 
\begin{equation}
\mathcal{G}(\eta)=\mathcal{G}_1(\eta)+\mathcal{G}_2(\eta)
\end{equation}
where 
\begin{eqnarray}
\mathcal{G}_1(\eta)&=&\int_{C_1}\frac{d \zeta}{2\pi i} \frac{\mathcal{G}(\zeta)}{\zeta-\eta}=\int\limits_1^\infty\frac{d \zeta}{2\pi i} \frac{{\rm Disc}_1[\mathcal{G}(\zeta)]}{\zeta-\eta} \;,\\
 \mathcal{G}_2(\eta)&=&-\int_{C_2}\frac{d \zeta}{2\pi i} \frac{\mathcal{G}(\zeta)}{\zeta-\eta}=\int\limits_{-\infty}^0\frac{d \zeta}{2\pi i} \frac{{\rm Disc}_2[\mathcal{G}(\zeta)]}{\zeta-\eta}\;,
\end{eqnarray}
and
\begin{eqnarray}
{\rm Disc}_1[\mathcal{G}(\zeta)]&=&\mathcal{G}(\zeta+i0^+)-\mathcal{G}(\zeta-i0^+)\;,\quad \text{for}\;\zeta\in(1,\infty)\;,\\
{\rm Disc}_2[\mathcal{G}(\zeta)]&=&\mathcal{G}(\zeta+i0^+)-\mathcal{G}(\zeta-i0^+)\;,\quad \text{for}\;\zeta\in(-\infty,0)\;.
\end{eqnarray}
Let us define 
\begin{equation}
k_h(\eta)=\eta^h{}_2F_1(h,h;2h,\eta)\;.
\end{equation}
These functions satisfy the following orthonormality condition
\begin{equation}\label{orthonormalk}
\oint_{|\eta|=\epsilon} \frac{d\eta}{2\pi i}\eta^{-2}k_{x+n}(\eta)k_{1-x-m}(\eta)=\delta_{mn}\;.
\end{equation}
Note that for $d=2$ and $\Delta_1=\Delta_2=\Delta_\phi$ the double-trace conformal blocks are just 
\begin{equation}
g_{\Delta_n^{d.t.}}(\eta)=\eta^{-\Delta_\phi}k_{\Delta_\phi+n}(\eta)\;.
\end{equation}
We expect that the Cauchy kernel can be expanded in terms of the double-trace conformal blocks 
\begin{equation}
\frac{1}{\zeta-\eta}=\sum_{n=0}^\infty H_n(\zeta)g_{\Delta_n^{d.t.}}(\eta)\;.
\end{equation}
The coefficients $H_n(\zeta)$ can then be extracted using the orthonormality relation (\ref{orthonormalk}) 
\begin{equation}
H_n(\zeta)=\oint_{|\eta|=\epsilon}\frac{d\eta}{2\pi i}\frac{\eta^{\Delta_\phi-2}}{\zeta-\eta}k_{1-\Delta_\phi-n}(\eta)\;.
\end{equation}
The integral is simple to perform and gives 
\begin{equation}
H_n(\zeta)=\frac{(-4)^{-n}(\Delta_\phi)_n(2\Delta_\phi-1)_n}{n!(\Delta_\phi-\frac{1}{2})_n}\zeta^{-1} {}_3F_2(1,-n,2\Delta_\phi+n-1;\Delta_\phi,\Delta_\phi,\zeta^{-1})\;.
\end{equation}
Therefore we have shown that $\mathcal{G}_1(\eta)$ can be expanded in terms of the bulk channel double-trace conformal blocks 
\begin{equation}
\mathcal{G}_1(\eta)=\sum_{n=0}^\infty r_{n,1}\, g_{\Delta_n^{d.t.}}(\eta)
\end{equation}
where 
\begin{equation}\label{defcn1}
r_{n,1}=\int_{C_1} \frac{d\zeta}{2\pi i} H_n(\zeta)\mathcal{G}(\zeta)\;.
\end{equation}
We can perform a similar analysis for $\mathcal{G}_2$. However, by crossing symmetry 
\begin{equation}
\mathcal{G}_2(\eta)=\pm \mathcal{G}_1(1-\eta)
\end{equation}
where $\pm$ is common parity of the two operators. It follows that $\mathcal{G}_2(\eta)$ admits an expansion into mirror channel double-trace conformal blocks
\begin{equation}
\mathcal{G}_2(\eta)=\sum_{n=0}^\infty r_{n,2}\, \bar{g}_{\Delta_n^{d.t.}}(\eta)
\end{equation}
where 
\begin{equation}\label{defcn2}
r_{n,2}=-\int_{C_2}\frac{d\zeta}{2\pi i}H_n(1-\zeta)\mathcal{G}(\zeta)\;.
\end{equation}
In fact the above decomposition is even valid when we do not assume crossing symmetry. To see this, we simply need to notice 
\begin{equation}
\frac{1}{\zeta-\eta}=-\frac{1}{(1-\zeta)-(1-\eta)}=-\sum_{n=0}^\infty H_n(1-\zeta)\bar{g}_{\Delta_n^{d.t.}}(\eta)\;.
\end{equation}

We have now established that the double-trace conformal blocks $g_{\Delta_n^{d.t.}}$ and $\bar{g}_{\Delta_n^{d.t.}}$ form a basis for functions satisfying the Regge behavior (\ref{spaceU}), which is captured by the decomposition 
\begin{equation}
\mathcal{G}(\eta)=\sum_{n=0}^\infty r_{n,1}\, g_{\Delta_n^{d.t.}}(\eta)+\sum_{n=0}^\infty r_{n,2}\, \bar{g}_{\Delta_n^{d.t.}}(\eta)\;.
\end{equation}
The coefficients can then be interpreted as the actions of the dual functionals 
\begin{equation} \label{FunctionalActionCorrelator}
r_{n,1}=\omega_n[\mathcal{G}(\eta)]\;,\quad r_{n,2}=\bar{\omega}_n[\mathcal{G}(\eta)]\;.
\end{equation}
Using their definitions (\ref{defcn1}), (\ref{defcn2}), we see that the functional kernels indeed follow from the dispersion relation kernel as we claimed earlier. 

The discussion in this section is quite similar to the CFT$_1$ case discussed in Section 2 of \cite{Mazac:2019shk}. But we should notice that the basis in Section 2 of \cite{Mazac:2019shk} does not coincide with the expected basis from holography. Holography suggests a basis containing all conformal blocks with dimensions $2\Delta_\phi+2n$ and their derivatives with respect to the conformal dimension, while the basis from the Cauchy dispersion kernel consists of all conformal blocks with dimensions $2\Delta_\phi+n$ and no derivatives. This is different in the $\mathbb{RP}^d$ CFT case. We find that the Cauchy dispersion kernel gives exactly the same double-trace basis which we expect from holography.

\subsubsection*{Nonperturbative checks}
Let us now perform some quick checks of the equivalence between functional actions obtained from the dispersion relation, and the ones obtained from Witten diagrams. Consider the following crossing symmetric toy example of a correlation function 
\begin{equation}
\mathcal{G}(\eta) = \frac{1}{\sqrt{\eta (1 - \eta)}}.
\end{equation}
We will set the external dimensions to be $\Delta_1 = \Delta_2 = 2$. The `correlator' can be decomposed into the $d = 2$ conformal blocks with a spectrum $\Delta=1+2n$, $n\in\mathbb{Z}_+$
\begin{equation}
\begin{split}
\mathcal{G}(\eta) &= \sum_{n = 1}^{\infty} \alpha_n g_{\Delta = 1 + 2 n}(\eta)\\
\alpha_n  &= \frac{(-1)^n 2^{1-2 n} \left(n-\frac{1}{2}\right)! \left(2 (-1)^n \Phi \left(-1,1,n+\frac{1}{2}\right)-\pi \right)}{\sqrt{\pi } (n-1)!}
\end{split}
\end{equation}
where $\Phi$ here is the Lerch transcendent function. Note that the mean field theory double-trace operators have conformal dimensions $\Delta=4+2n$, $n\in \mathbb{N}$. The OPE spectrum of the above correlator is therefore `maximally' different from the mean field spectrum, and is in this sense a `nonperturbative' check. We can act with the functionals on both sides using \eqref{FunctionalActionBlockDirect} and \eqref{FunctionalActionCorrelator}, and it leads to the following constraints
\begin{equation}
\omega_{n} [\mathcal{G}(\eta)] = \int_1^{\infty} \frac{d \zeta}{2 \pi i} {\rm Disc}_1 [\mathcal{G}(\zeta) H_{n} (\zeta)] = -\sum_{m = 1}^{\infty} \alpha_m  \left( \frac{A_n}{A} \right)_{\Delta= 1 + 2 m, \ \Delta_1 = \Delta_2 = 2}.
\end{equation}
We checked numerically that this constraint is true for $n = 0, 1$ and $2$. 

A more physical example is given by the $2 d$ Ising model. The exact solution is known for this model. The $\sigma$ operator, which has conformal dimension $\Delta_{\sigma} = 1/8$, has a two point function on $\mathbb{RP}^d$ given by (we assume that $\sigma$ has positive parity) \cite{Nakayama:2016cim}
\begin{equation}
\mathcal{G}_{\sigma}(\eta) = \frac{\sqrt{1-\sqrt{1-\eta }}+\sqrt{1-\sqrt{\eta }}}{(\eta  (1-\eta ))^{1/8}}  = \sum_{n = 0}^{\infty} \rho_{1,n} g_{\Delta = 4 n}(\eta) + \sum_{n = 1}^{\infty} \rho_{2,n} g_{\Delta = 1 + 4 n}(\eta).
\end{equation}
The coefficients $\rho_{1,n}$ and $\rho_{2,n}$ can be found recursively using the expansion of the correlator. Note that $\mathcal{G}_{\sigma}(\eta)$ approaches a constant as $\eta\to \infty$, and therefore does not belong to the space of functions we defined. In fact, a direct application of the functionals leads to a divergent sum. So instead we perform check on $\mathcal{G}_{\sigma}(\eta)/ \eta $ which has an improved Regge behavior. We would like the new correlator to have the same operator spectrum in the conformal block decomposition. Therefore, we will take the external dimensions to be $\Delta_1 = \Delta_2 = 9/8$ instead of $1/8$. The action of the functionals then requires 
\begin{equation}
\begin{split}
\omega_{n} \left[\frac{\mathcal{G}_{\sigma}(\eta)}{\eta}\right] &= \int_1^{\infty} \frac{d \zeta}{2 \pi i} {\rm Disc}_1 \left[\frac{\mathcal{G}_{\sigma}(\zeta)}{\zeta} H_{n} (\zeta)\right] \\
&= -\sum_{m = 0}^{\infty} \rho_{1,m}  \left( \frac{A_n}{A} \right)_{\Delta= 4 m, \ \Delta_1 = \Delta_2 = \frac{9}{8}} -\sum_{m = 0}^{\infty} \rho_{2,m}  \left( \frac{A_n}{A} \right)_{\Delta= 4 m + 1, \ \Delta_1 = \Delta_2 = \frac{9}{8}} . 
\end{split}
\end{equation}
We checked this relation numerically for $n = 0, 1$ and $2$ and it holds true.

\subsection{Perturbative applications of analytic functionals}\label{Sec:testfunctionals}
In this subsection, we apply our functionals to some perturbative examples. We start by checking the sum rules \eqref{sumrulea}, \eqref{sumruleb} on the mean field theory. We then consider small perturbations around the mean field theory and show how we can obtain the data for Wilson-Fisher model on $\mathbb{RP}^d$ using these sum rules. Note that the mean field theory two-point function has the following conformal block decomposition 
\begin{equation}\label{CBDGFF}
\mathcal{G}(\eta) = \frac{1}{\eta^{\Delta_{\phi}}} \pm \frac{1}{(1 - \eta)^{\Delta_{\phi}}} = g_{\Delta = 0}(\eta) + \sum_{n = 0}^{\infty} \mu_{\phi \phi n} g_{\Delta = \Delta_n^{d.t.}} (\eta)
\end{equation}
where $\Delta_n^{d.t.} = 2 \Delta_{\phi} + 2 n$ and the OPE coefficients are given by 
\begin{equation}\label{GFFCoefficients}
\mu_{\phi \phi n}^{(0)} = \pm \frac{(\Delta_{\phi})_n (2 \Delta_{\phi} - \frac{d}{2} + 2 n)_{-n}}{(\Delta_{\phi} - \frac{d}{2} + n + 1)_{-n} n!}.
\end{equation}
These OPE coefficients are just a simple modification of the BCFT case, which can be found for instance in \cite{Liendo:2012hy}. We use the superscript $0$ to indicate that we will soon perturb the mean field theory solution. The sum rules then tells us that the following must be true for all values of $n$ 
\begin{equation} \label{ConstraintGFF}
- \left( \frac{A_n}{A}\right)_{\Delta = 0}  + \mu_{\phi \phi n}^{(0)} = \pm \left( \frac{B_n}{A} \right)_{\Delta = 0}.
\end{equation}
Now recall the expression of the coefficients $A_{n}$ and $A$ from \eqref{WAcoe} and \eqref{WAnDDT}. For $\Delta_1 = \Delta_2 = \Delta_{\phi}$ they take a simpler form
\begin{equation} \label{WCoefGFF}
\begin{split}
A_{n} &= \frac{(- 1)^{n } (\Delta_{\phi})_n^2 (2 \Delta_{\phi} - \frac{d}{2} + 2 n)_{-n}}{\left(\Delta(\Delta - d) - (2 \Delta_{\phi} + 2 n)(2 \Delta_{\phi} + 2 n - d) \right) n!} \\
A &= \frac{\Gamma\left( \frac{\Delta}{2} \right)^2 \Gamma\left( \Delta_{\phi} - \frac{\Delta}{2} \right) \Gamma\left( \Delta_{\phi} - \frac{(d - \Delta)}{2}  \right)}{4 \Gamma(\Delta_{\phi})^2 \Gamma\left( \Delta + 1 - \frac{d}{2} \right)}.
\end{split}
\end{equation}
It is then clear that as we take $\Delta \rightarrow 0$ the expression for the coefficient $A$ diverges, while $A_m$ remains finite. So the first term in the constraint equation \eqref{ConstraintGFF}, $A_m/A$ does not contribute. The constraint then becomes
\begin{equation} \label{BFunctionalIdentity}
\left( \frac{B_n}{A} \right)_{\Delta = 0} = \pm \mu_{\phi \phi n}^{(0)}
\end{equation}
This constraint can be explicitly checked for $n = 0$ and $n = 1$ using the results for $B_n$ in \eqref{B0BoundaryCondition} and \eqref{BnRecursion}. For all other values of $n$, note that the recursion relation \eqref{BnRecursion} implies 
\begin{equation}
\begin{split}
&\left( \frac{\rho_{n - 1} B_{n - 1}}{A} \right)_{\Delta = 0} + \left( \frac{\nu_n B_n}{A} \right)_{\Delta = 0} + \left( \frac{\mu_{n + 1} B_{n + 1}}{A} \right)_{\Delta = 0} = \left( \frac{a_n}{A} \right)_{\Delta = 0} = 0 
\\
\\
& \implies \left( \rho_{n - 1} \ \mu_{\phi \phi n-1}^{(0)} \ + \ \nu_n \ \mu_{\phi \phi n}^{(0)} \ + \ \nu_n \ \mu_{\phi \phi n}^{(0)}  \right)_{\Delta = 0} = 0
\end{split}
\end{equation}
which can be easily checked to be true for all values of $n$. This completes our check of analytic functionals for mean field theory. 

\subsubsection*{Wilson-Fisher model}
We now consider perturbations around the mean field solution such that the above OPE coefficients $\mu_{\phi \phi n}$ and dimensions receive small corrections. One such perturbation is the Wilson-Fisher fixed point in $d = 4 - \epsilon$ which is a perturbation of free field theory with $\Delta^{(0)}_{\phi} = \frac{d}{2} - 1$. It has a Lagrangian description, which in our normalization can be written as 
\begin{equation}
S = \frac{\Gamma \left(\frac{d}{2} - 1 \right)}{4 \pi^{d/2}} \int d^d x \left( \frac{1}{2} (\partial_{\mu} \phi^I)^2 + \frac{\lambda}{4} (\phi^I \phi^I)^2 \right). 
\end{equation}
But we will not need this Lagrangian description, and we will treat it as a perturbation of a mean field theory of $N$ free fields. We parametrize deviations from the mean field values as follows
\begin{equation}
\begin{split}
\mu_{\phi \phi n} &= \mu_{\phi \phi n}^{(0)} + \epsilon \ \mu_{\phi \phi n}^{(1)} + \epsilon^2 \ \mu_{\phi \phi n}^{(2)}, \hspace{1cm}  \Delta_{\phi} = \frac{d}{2} - 1 + \epsilon^2 \gamma^{(2)}_{\phi} \\
\Delta & = \Delta_n^{d.t.} + \epsilon \gamma_n^{(1)} + \epsilon^2 \gamma_n^{(2)} = 2 \Delta_{\phi} + 2 n + \epsilon \gamma_n^{(1)} + \epsilon^2 \gamma_n^{(2)}
\end{split}
\end{equation}
where we used the well known fact that in this model, the first order correction to the anomalous dimension of $\phi$ vanishes. From \eqref{GFFCoefficients}, we see that for this free field value of $\Delta^{(0)}_{\phi}$, $\mu_{\phi \phi n}^{(0)} = \pm \delta_{n , 0}$, which truncates the functional equations. This leaves us with a finite number of terms on both sides. Using \eqref{orthonormal} the sum rule \eqref{sumrulea}  at order $\epsilon$ just becomes 
\begin{equation}
\mu_{\phi \phi n}^{(1)} \mp \left( \frac{A_n}{A}\right)^{O(\epsilon)}_{\Delta = \Delta_{0}^{d.t.} \ + \ \epsilon \gamma_0^{(1)}} = \pm\left( \frac{B_n}{A}\right)^{O(\epsilon)}_{\Delta = 0} +  \left( \frac{B_n}{A}\right)^{O(\epsilon)}_{\Delta = \Delta_0^{d.t.} \ + \  \epsilon \gamma_0^{(1)}} 
\end{equation}
and the superscript indicates that we pick out the order $\epsilon$ contribution. Expanding in $\epsilon$, we can check that $(A_n/A)$ term does not contribute at order $\epsilon$. Also using \eqref{BFunctionalIdentity} and \eqref{GFFCoefficients}, we can check that $(B_n/A)$ does not contribute for $\Delta = 0$. As for the other term on the right hand side, it only contributes at this order for $n = 0$ and $1$ and using the recursion relation \eqref{BnRecursion}, we can check that all the other values of $n$ start contributing at order $\epsilon^2$. This gives the following results for the CFT data
\begin{equation}
\mu_{\phi \phi 0}^{(1)} = - \frac{\gamma^{(1)}_0}{2}, \ \ \ \mu_{\phi \phi 1}^{(1)} =  \frac{\gamma^{(1)}_0}{4}, \ \ \ \mu_{\phi \phi n \ge 2}^{(1)} = 0 
\end{equation}
This agrees with what was found in \cite{Hasegawa:2018yqg}. The fact that only two of the OPE coefficients are non-zero at this order implies that the functional equations also truncate to finite terms at next order. At next order in $\epsilon$, we obtain using the sum rule
\begin{equation}
\begin{split} \label{Epsilon2OPESumRule}
&\mu_{\phi \phi n}^{(2)} \mp \left( \frac{A_n}{A}\right)^{O(\epsilon^2)}_{\Delta = \Delta_{0}^{d.t.}  +  \epsilon \gamma_0^{(1)} +  \epsilon^2 \gamma_0^{(2)}} -  \mu_{\phi \phi 0}^{(1)} \left( \frac{A_n}{A}\right)^{O(\epsilon)}_{\Delta = \Delta_{0}^{d.t.} + \epsilon \gamma_0^{(1)}} -  \mu_{\phi \phi 1}^{(1)} \left( \frac{A_n}{A}\right)^{O(\epsilon)}_{\Delta = \Delta_{1}^{d.t.} + \epsilon \gamma_1^{(1)}} = \\
&\pm \left( \frac{B_n}{A}\right)^{O(\epsilon^2)}_{\Delta = 0} +  \left( \frac{B_n}{A}\right)^{O(\epsilon^2)}_{\Delta = \Delta_{0}^{d.t.} + \epsilon \gamma_0^{(1)}  +  \epsilon^2 \gamma_0^{(2)}}  \pm \mu_{\phi \phi 0}^{(1)} \left( \frac{B_n}{A}\right)^{O(\epsilon)}_{\Delta = \Delta_{0}^{d.t.} + \epsilon \gamma_0^{(1)}} \pm  \mu_{\phi \phi 1}^{(1)} \left( \frac{B_n}{A}\right)^{O(\epsilon)}_{\Delta = \Delta_{1}^{d.t.} +  \epsilon \gamma_1^{(1)}}.
\end{split}
\end{equation}
For the identity block, using \eqref{BFunctionalIdentity} and \eqref{GFFCoefficients}, it is easy to check that 
\begin{equation}
\left( \frac{B_n}{A}\right)^{O(\epsilon^2)}_{\Delta = 0}  = \frac{\gamma_{\phi}^{(2)} (\Gamma(n))^2}{\Gamma(2 n)}, n \ge 0 \hspace{1 cm} \left( \frac{B_0}{A}\right)^{O(\epsilon^2)}_{\Delta = 0}  = 0.
\end{equation}
For other values of $\Delta$, expanding the $A_n$-functionals is straightforward. To expand the $B_n$-functionals, which involve hypergeometric functions, we use the package $\mathtt{HypExp}$ \cite{Huber:2005yg}. We collect below the needed expansions for a few low-lying values of $n$
\begin{equation}
\begin{split}
\left( \frac{B_0}{A}\right)_{\Delta = \Delta_{0}^{d.t.} + \epsilon \gamma_0^{(1)} +  \epsilon^2 \gamma_0^{(2)}} &=  - \frac{\gamma_0^{(1)}}{2} \epsilon -  \frac{\left(\gamma_0^{(1)} + 2 \gamma_0^{(2)}\right)}{4} \epsilon^2; \ \left( \frac{B_0}{A}\right)_{\Delta = \Delta_{1}^{d.t.} + \epsilon \gamma_1^{(1)}} = \frac{\gamma_1^{(1)} (\pi^2 - 6)}{6} \epsilon, \\
\left( \frac{B_1}{A}\right)_{\Delta = \Delta_{0}^{d.t.} +  \epsilon \gamma_0^{(1)} +  \epsilon^2 \gamma_0^{(2)}} &=   \frac{\gamma_0^{(1)}}{4} \epsilon +  \frac{\left( \gamma_0^{(1)} (2\gamma_0^{(1)} - 1 ) + 4 \gamma_0^{(2)}\right)}{16} \epsilon^2; \ \left( \frac{B_1}{A}\right)_{\Delta = \Delta_{1}^{d.t.} + \epsilon \gamma_1^{(1)}} = -  \frac{\gamma_1^{(1)}}{2} \epsilon, \\
\left( \frac{B_2}{A}\right)_{\Delta = \Delta_{0}^{d.t.} +  \epsilon \gamma_0^{(1)} +  \epsilon^2 \gamma_0^{(2)}} &= -  \frac{\gamma_0^{(1)} (\gamma_0^{(1)} - 1)}{24} \epsilon^2; \ \left( \frac{B_2}{A}\right)_{\Delta = \Delta_{1}^{d.t.} + \epsilon \gamma_1^{(1)}} = \frac{\gamma_1^{(1)}}{12} \epsilon, \\
\left( \frac{B_3}{A}\right)_{\Delta = \Delta_{0}^{d.t.}+  \epsilon \gamma_0^{(1)} +  \epsilon^2 \gamma_0^{(2)}} &=  \frac{\gamma_0^{(1)} (\gamma_0^{(1)} - 1)}{480} \epsilon^2; \ \left( \frac{B_3}{A}\right)_{\Delta = \Delta_{1}^{d.t.} + \epsilon \gamma_1^{(1)}} = -\frac{\gamma_1^{(1)}}{90} \epsilon.
\end{split}
\end{equation}
Going to higher values of $n$ is also straightforward by using the recursion relation \eqref{BnRecursion}. Using these expansions of coefficients, we can obtain the results of $\mu_{\phi \phi n}^{(2)}$ with $n=0,1,2,3$  
\begin{equation}
\begin{split}
\mu_{\phi \phi 0}^{(2)} &= \frac{\gamma_0^{(1)} (\gamma_1^{(1)} - 1)}{4} - \frac{\gamma_0^{(2)}}{2} \pm \frac{\gamma_0^{(1)} (\gamma_0^{(1)} - \gamma_1^{(1)})}{4}\\
\mu_{\phi \phi 1}^{(2)} &= \frac{\gamma_0^{(1)} (2 \gamma_0^{(1)} - 3 \gamma_1^{(1)} - 1)}{16} + \frac{\gamma_0^{(2)}}{4} \mp \frac{\gamma_0^{(1)} (3 \gamma_0^{(1)} + \gamma_1^{(1)} - 2)}{8} \pm \gamma_{\phi}^{(2)} \\
\mu_{\phi \phi 2}^{(2)} &= \pm \frac{ \gamma_0^{(1)}(\gamma_0^{(1)} - 1 + \gamma_1^{(1)})}{48} - \frac{ \gamma_0^{(1)}(\gamma_0^{(1)} - 1)}{24} - \frac{ \gamma_0^{(1)}\gamma_1^{(1)}}{36} \pm \frac{\gamma_{\phi}^{(2)}}{6} \\
\mu_{\phi \phi 3}^{(2)} &= \frac{\gamma_0^{(1)} (\gamma_0^{(1)} - 1)}{480} + \frac{\gamma_0^{(1)} \gamma_1^{(1)}}{320} \mp \frac{\gamma_0^{(1)}(\gamma_0^{(1)} - 1 + \gamma_1^{(1)})}{360} \pm \frac{\gamma_{\phi}^{(2)}}{30}.
\end{split}
\end{equation}
The bulk data of the $O(N)$ vector model at Wilson-Fisher fixed point can be found in \cite{PhysRevD.7.2911, Rychkov:2015naa, Gliozzi:2017hni}
\begin{equation}
\gamma_0^{(1)} = \frac{N + 2}{N + 8}, \hspace{1cm} \gamma_1^{(1)} = 1, \hspace{1cm}  \gamma_0^{(2)} =  \frac{6 (N + 2) (N + 3)}{(N + 8)^3}, \hspace{1cm} \gamma_{\phi}^{(2)} = \frac{N + 2}{4 (N + 8)^2}.
\end{equation} 
This then gives us the following $\mathbb{RP}^d$ OPE coefficients  to the $\epsilon^2$ order
\begin{equation} \label{Epsilon2OPEResults}
\begin{split}
\mu_{\phi \phi 0} &= \pm 1 - \frac{N + 2}{2 (N + 8)} \epsilon  - \frac{3 (N + 2) (2 N + 6 \pm (N + 8))}{2 (N + 8)^3} \epsilon^2  \\
\mu_{\phi \phi 1} &=  \frac{N + 2}{4 (N + 8)} \epsilon  - \frac{ (N + 2) }{4 (N + 8)^2} \left( \frac{76 + N(N + 10)}{2 (N + 8)} \pm (N - 2)  \right) \epsilon^2  \\
\mu_{\phi \phi 2} &= \frac{N + 2}{48 (N + 8)^2} \left( \pm (N + 4) - \frac{4}{3} (N - 1) \right) \epsilon^2 \\
\mu_{\phi \phi 3} &= \frac{N + 2}{320 (N + 8)^2} \left( (N + 4) \mp \frac{8}{9} (N + 2) \pm \frac{8}{3} \right) \epsilon^2.
\end{split}
\end{equation}   
We can obtain  numerical estimations for OPE coefficients in the $d = 3$ Ising model by  plugging in $\epsilon = 1$ and $N = 1$ in the above expressions. This gives in particular $\mu^{+}_{\phi \phi 0} = 0.728$ and $\mu^{+}_{\phi \phi 1} = 0.0478$, which can be compared with  the results obtained by the bootstrap analysis  \cite{Nakayama:2016cim}. The bootstrap result gives $\mu^{+}_{\phi \phi 0} \sim 0.70$ and $\mu^{+}_{\phi \phi 1} \sim 0.047$, and is in good agreement with our result. We can keep going and it is completely straightforward to obtain all the OPE coefficients at order $\epsilon^2$ using \eqref{Epsilon2OPESumRule}. On the other hand, since all the OPE coefficients are non-zero at this order, the sum rules at the next order will contain infinite number of terms. The sum rules still put nontrivial constraints on the OPE coefficients, but it is not clear how the constraints can be solved analytically. 

\subsubsection*{Large N checks}
We now provide an independent consistency check of some of these results by considering the large $N$ $O(N)$ vector model and compare the results in that with the large $N$ limit of \eqref{Epsilon2OPEResults}.

Let us first note that to the leading order in $\epsilon$, we can use the results from \eqref{Epsilon2OPEResults} to write the two-point function as 
\begin{equation}
\langle \phi^I (x_1) \phi^J (x_2) \rangle  = \frac{\delta^{IJ} \mathcal{G}(\eta)}{((1 + x_1^2)(1 + x_2^2))^{\Delta_{\phi}}}
\end{equation}
with 
\begin{equation} \label{TwoPointEpsilon1}
\begin{split}
\mathcal{G}(\eta) &= g_{\Delta = 0} (\eta) + \mu_{\phi \phi 0} g_{\Delta = \Delta_0^{d.t.}} (\eta) + \mu_{\phi \phi 1} g_{\Delta = \Delta_1^{d.t.}} (\eta) \\
&= \frac{1}{\eta^{\frac{d}{2} - 1}} \left( 1 \pm  \left(\frac{\eta}{1 - \eta} \right)^{\frac{d}{2} - 1} \pm \frac{\epsilon (N + 2)}{2 (N + 8)} \left(\frac{\eta}{1 - \eta} \right) \log \eta +  \frac{\epsilon (N + 2)}{2(N + 8)} \log (1 - \eta)   \right)  
\end{split}
\end{equation}

To develop a large $N$ expansion, we introduce the usual Hubbard-Stratonovich auxiliary field $\sigma$ and write down the action as 
\begin{equation}
S = \frac{\Gamma \left(\frac{d}{2} - 1 \right)}{4 \pi^{d/2}} \int d^d x \left( \frac{1}{2} (\partial_{\mu} \phi^I)^2 + \frac{\sigma \phi^I \phi^I}{2} \right)
\end{equation}
where we omitted a $\sigma^2/ 4 \lambda$ term, which can be dropped at the fixed point. At leading order in large $N$, we get the following equation of motion for the $\phi$ correlator 
\begin{equation}
(\nabla^2 - \langle \sigma (x_1)\rangle  ) \langle \phi^I (x_1) \phi^J (x_2) \rangle = - \frac{4 \pi^{d/2}}{\Gamma \left( \frac{d}{2} - 1 \right)}\delta^{I J} \delta^d (x_1 - x_2).
\end{equation} 
At large $N$ the scaling dimension of $\sigma$ is $2 + 1/N$, while the scaling dimension of $\phi$ is $d/2 - 1 + 1/N$. We can again express these correlators as
\begin{equation} \label{LargeNCorrParamet}
\langle \sigma (x_1)\rangle = \frac{a_{\sigma}}{(1 + x_1^2)^2}, \hspace{1 cm} \langle \phi^I (x_1) \phi^J (x_2) \rangle  = \frac{\delta^{IJ} \mathcal{G}(\eta)}{((1 + x_1^2)(1 + x_2^2))^{\Delta_{\phi}}} = \frac{\delta^{IJ} G(\eta)}{((x_1 - x_2)^2)^{\Delta_{\phi}}}.
\end{equation}
Plugging in this general form into the equation of motion, we get 
\begin{equation}
4 \eta (1 - \eta) \frac{d  G (\eta)}{d \eta^2} + (8 (1 - \eta) -2 d) \frac{d G (\eta) }{d \eta} - a_{\sigma} G (\eta) = 0.
\end{equation}
This equation can be solved and the general solution is
\begin{equation} \label{LargeNTwoPointSol}
\begin{split}
G(\eta) =&   C_1 \left( \frac{\eta}{1 - \eta} \right)^{\frac{d}{2} - 1} \ {}_2F_1 \left( \frac{1 + \sqrt{1 - a_{\sigma}}}{2}, \frac{1 - \sqrt{1 - a_{\sigma}}}{2},  2 - \frac{d}{2}, 1 - \eta \right) \\
&+  C_2 \ {}_2F_1 \left( \frac{1 + \sqrt{1 - a_{\sigma}}}{2}, \frac{1 - \sqrt{1 - a_{\sigma}}}{2},  \frac{d}{2}, 1 - \eta \right).
\end{split}
\end{equation}
Recall that the crossing equation \eqref{crossingeqn} requires 
\begin{equation}
\frac{G(\eta)}{\eta^{\frac{d}{2} - 1}} = \pm \frac{G(1 - \eta)}{(1 - \eta)^{\frac{d}{2} - 1}} 
\end{equation}
which implies that the coefficients must satisfy
\begin{equation} \label{LargeNTwoPointCoeff}
\frac{C_2}{C_1} = \pm \frac{\Gamma \left( \frac{d - 1 + \sqrt{1 - a_{\sigma}}}{2} \right)\Gamma \left( \frac{d - 1 - \sqrt{1 - a_{\sigma}}}{2} \right) \Gamma\left(2 - \frac{d}{2} \right)}{\pi \Gamma\left( \frac{d}{2} \right)} \left( - \sin \pi \left( \frac{d \pi }{2}\right) \mp \cos \left( \frac{\pi  \sqrt{1 - a_{\sigma}}}{2} \right)\right). 
\end{equation} 
The overall constant can then be fixed by demanding that the leading term in small $\eta$ expansion of $G$ is just 1. We can expand this solution in small $\eta$ as
\begin{equation}
\begin{split}
&G(\eta) = C_1 \left( \pm 1 \ \mp \ \frac{a_{\sigma} }{2 (d - 4) } \ \eta + O(\eta^2) \right)   \\
& + C_1 \eta^{\frac{d}{2} - 1} \left(  \frac{\pi \Gamma \left( 2 - \frac{d}{2} \right)}{ \left(\sin \left( \frac{ \pi d}{2} \right)  \mp \cos \left( \frac{\pi  \sqrt{1 - a_{\sigma}}}{2} \right) \right)  
\Gamma \left( \frac{d}{2} \right) \Gamma\left( 
\frac{3 - d - \sqrt{1 - a_{\sigma}}}{2} \right) \Gamma \left( \frac{3 - d + \sqrt{1 - a_{\sigma}}}{2} \right)} + O(\eta) \right).
\end{split}
\end{equation}
The first term in the first line is the contribution from the identity operator, while the second term of order $\eta$ is from the $\sigma$ operator of dimension $2$. The term in the second line represents the $\phi^2$ operator of dimension $d - 2$. In the large $N$ theory, we expect the $\phi^2$ operator of dimension $d - 2$ in the free theory to be replaced by the $\sigma$ operator of dimension $2$. This then requires us to set the term in the second line of the above equation to zero, which implies the following possible values of $a_{\sigma}$
\begin{equation}
 a^{+}_{\sigma} = - (d - 2) (d - 4), - (d - 6)(d - 8), .... \hspace{1cm} a^{-}_{\sigma} = - (d - 4)(d - 6), - (d - 8)(d - 10), ....
\end{equation}
Note that for these values of $a_{\sigma}$, the coefficient $C_{2}$ also vanishes. Now we expect the large $N$ theory to match with the free theory in $d = 4$. This means that we must choose the value of $a_{\sigma}$ such that $a_{\sigma} = 0$ as $d \rightarrow 4$. This then picks out the solution for us for both $+$ and $-$ parity \footnote{In appendix \ref{AppendixFreeEnergy}, we provide another way to check this value of $a_{\sigma}$, where it occurs as a large $N$ saddle point of the free energy.}
\begin{equation} \label{TwoPointCriticalO(N)+-}
\begin{split}
&a^+_{\sigma} = - (d - 2) (d - 4), \implies G^+ (\eta) = \frac{1}{(1 - \eta)^{\frac{d}{2} - 1}}\\
&a^-_{\sigma} = - (d - 4) (d - 6) \implies G^-(\eta) = \frac{1 - 2 \eta}{(1 - \eta)^{\frac{d}{2} - 1}}.
\end{split}
\end{equation}
It can be checked that these results in $d = 4- \epsilon$ agree with the large $N$ limit of the $\epsilon$ expansion solution \eqref{TwoPointEpsilon1}. These two point functions can be decomposed into conformal blocks of dimension $2 n + 2$ as follows 
\begin{equation} \label{CBDLargeN-}
\begin{split}
\mathcal{G}^-(\eta) &= \frac{1}{(\eta(1 - \eta))^{\frac{d}{2} - 1}} = g_{\Delta = 0}(\eta) + \sum_{n = 0}^{\infty} \lambda^+_n g_{\Delta = 2 n + 2} (\eta)  \\
\lambda^+_n &= \frac{\Gamma \left(\frac{d}{2}\right) \, _2F_1\left(-n-1,-n;\frac{1}{2} (d-4 n-2);1\right)}{\Gamma (n+2) \Gamma \left(\frac{d}{2}-n-1\right)} 
\end{split}
\end{equation}
for the $+$ case and 
\begin{equation} \label{CBDLargeN+}
\begin{split}
\mathcal{G}^-(\eta) &= \frac{1 - 2 \eta}{(\eta(1 - \eta))^{\frac{d}{2} - 1}} = g_{\Delta = 0}(\eta) + \sum_{n = 0}^{\infty} \lambda^-_n g_{\Delta = 2 n + 2} (\eta)  \\
\lambda^-_n &= -\frac{ (-1)^n (d^2 - 4 d (n + 2) + 8 (1 + n)^2 + 4) \Gamma(1 - \frac{d}{2} + n) \Gamma(2 - \frac{d}{2} + n)^2}{4 \Gamma(2 - \frac{d}{2})^2 \Gamma(n + 2) \Gamma(2 - \frac{d}{2} + 2 n)}
\end{split}
\end{equation}
for the $-$ case. These coefficients can be found in a manner similar to the one used for BCFT case which can be found in \cite{Liendo:2012hy, Giombi:2020rmc}. We want to emphasize here that unlike the mean field theory, the conformal blocks appearing here have dimensions $2 n + 2$. These OPE coefficients can be expanded in $\epsilon$ in $d = 4 - \epsilon$, and we find a precise match with the large $N$ limit of \eqref{Epsilon2OPEResults}.

\subsection{Using CFT equations of motion}\label{Sec:CFTEoM}
A complementary approach to using the analytic functionals, where  a Lagrangian description for the CFT is available, is to use the CFT equations of motion.\footnote{Note that this is different from what we did in Section \ref{Sec:3}, where we used the equations of motion in the bulk $AdS_{d+1}/\mathbb{Z}_2$.} The essential idea was described for the CFT in flat space in \cite{Rychkov:2015naa} and extended to the case of BCFT in \cite{Giombi:2020rmc}. Here we will use this method to fix the two-point function of the field $\phi$ in the Wilson-Fisher model on $\mathbb{RP}^d$. This case is very similar to the BCFT case. Since the two-point function is only a function of cross-ratio $\eta$ on the sphere quotient $S^d/\mathbb{Z}_2$ and is proportional to $\mathcal{G}(\eta)$, in this subsection it will be more convenient to `undo' the Weyl transformation (\ref{MetricWeyl}) and work on the sphere quotient. 

The action for the $\phi^4$ Wilson-Fisher model, including the conformal coupling term, can be written as 
\begin{equation} \label{ActionSpherePhi4}
S = \frac{\Gamma \left(\frac{d}{2} - 1 \right)}{4 \pi^{d/2}} \int d^d x \left( \frac{1}{2} (\partial_{\mu} \phi^I)^2 + \frac{d (d - 2)}{4} \phi^I \phi^I + \frac{\lambda}{4} (\phi^I \phi^I)^2 \right). 
\end{equation}
The two-point function on the sphere is 
\begin{equation}
\langle \phi^I(x_1) \phi^J(x_2) \rangle = \frac{\delta^{IJ} \mathcal{G} (\eta)}{4}.
\end{equation}
Let us start with the free theory with $\lambda = 0$. The the field $\phi^I$ satisfies $(\nabla^2  - d (d - 2)/4) \phi^I = 0$, which implies that
\begin{equation}
\begin{split}
&\left( \frac{1}{\sqrt{g}} \partial_{\mu} (g^{\mu \nu} \sqrt{g} \partial_{\nu}) - \frac{d (d - 2)}{4}  \right) \mathcal{G}(\eta) = 0\\
&\eta (1 - \eta) \frac{d^2 \mathcal{G} }{d \eta^2} + d \left( \frac{1}{2} - \eta \right) \frac{d\mathcal{G} }{d \eta} - \frac{d (d - 2)}{4} \mathcal{G} = D^{(2)} \mathcal{G}(\eta) = 0.
\end{split}
\end{equation}
This equation can be solved, and the general solution is 
\begin{equation} \label{TwoPointSolFree}
\mathcal{G}(\eta) = b_1 \left( \frac{1}{\eta^{\frac{d}{2} - 1}}  + \frac{1}{(1 - \eta)^{\frac{d}{2} - 1}}\right) + b_2 \left( \frac{1}{\eta^{\frac{d}{2} - 1}}  - \frac{1}{(1 - \eta)^{\frac{d}{2} - 1}}\right).
\end{equation}
The constants are just fixed by the normalization, and we pick $b_1 = 1, b_2 = 0$ for $+$ parity, and vice versa for the $-$ parity. When we include interactions, the equation of motion gets modified to $(\nabla^2  - d (d - 2)/4) \phi^I(x) = \lambda \phi^I \phi^K \phi^K (x)$. This implies 
\begin{equation}
 D^{(2)} \mathcal{G}(\eta) =  \frac{\lambda_* (N + 2) a_{\phi^2}}{4} \mathcal{G}(\eta) + O(\lambda_*^2)
\end{equation}
to leading order in $\lambda$.  We can solve this perturbatively in $d = 4 - \epsilon$, where this model has a non-trivial fixed point. $\lambda_*$ is the fixed point value of the coupling and is equal to $2 \epsilon/ (N + 8)$ in this normalization, while $a_{\phi^2}/4$ is the one-point function of $\phi^2$ on the sphere. Since there is a factor of $\lambda_*$ on the right hand side, we can plug in the correlators in $d = 4$, and it will give us the two-point function on the left hand side, correct to order $\epsilon$. We can expand the differential operator and the correlator as follows
\begin{equation}
\begin{split}
\mathcal{G}(\eta) = \mathcal{G}_0(\eta) + \epsilon \mathcal{G}_1(\eta) + \epsilon^2 \mathcal{G}_2(\eta) + O(\epsilon^3) \\
D^{(2)} = D_0^{(2)} + \epsilon D_1^{(2)} + O(\epsilon^2)
\end{split}
\end{equation}
where $\mathcal{G}_0(\eta) $ is just given by \eqref{TwoPointSolFree} with $d = 4$. The equation of motion at first order in $\epsilon$ is 
\begin{equation}
D_0^{(2)} \mathcal{G}_1(\eta) = \pm \frac{N + 2}{2 (N + 8)} \mathcal{G}_0(\eta) - D_1^{(2)} \mathcal{G}_0(\eta). 
\end{equation}
This equation can also be solved te give
\begin{equation} \label{TwoPointSolEps}
\mathcal{G}_1(\eta) = \frac{c_1}{\eta} + \frac{c_2}{1 - \eta} +  \frac{\log \eta}{2 \eta} \pm \frac{\log( 1 - \eta)}{2 ( 1 - \eta)} + \frac{N + 2}{2 (N + 8)} \left( \frac{\log (1 - \eta)}{ \eta} \pm \frac{\log \eta}{1 - \eta}  \right).
\end{equation}
If we fix the normalization such that $\mathcal{G}(\eta) = \eta^{-1}$ as $\eta \rightarrow 0$, this fixes $c_1 = 0$. Also, the crossing symmetry \eqref{crossingeqn} requires $\mathcal{G}$ to be either symmetric or antisymmetric under $\eta \rightarrow 1 - \eta$, which sets $c_2 = 0$. This order $\epsilon$ correlator then agrees exactly with the result using functionals \eqref{TwoPointEpsilon1}. Now to go to the next order, note that in the two-point function $\phi^I (x_1) \phi^J(x_2)$, we can also apply the equation of motion to the other $\phi$. This gives the following fourth-order differential equation
\begin{equation}
D^{(4)}\mathcal{G}(\eta) =D^{(2)}( D^{(2)} \mathcal{G}(\eta)) = \frac{\lambda_*^2 (N + 2)}{16} (a_{\phi^2}^2 (N + 2)\mathcal{G}(\eta) + 2 \mathcal{G}(\eta)^3  ).
\end{equation}
We can again solve it perturbatively in $\epsilon$ by expanding $D^{(4)} = D_0^{(4)} +  \epsilon D_1^{(4)} + \epsilon^2 D_2^{(4)} + O(\epsilon^3)$. The differential equation at $O(\epsilon^2)$ then becomes
\begin{equation}
D_0^{(4)} \mathcal{G}_2(\eta) = \frac{(N + 2)}{4 (N + 8)^2} \left((N + 2)\mathcal{G}_0(\eta) + 2 \mathcal{G}_0(\eta)^3  \right) - D_1^{(4)} \mathcal{G}_1(\eta) - D_2^{(4)} \mathcal{G}_0(\eta). 
\end{equation}
This general solution of this equation is 
\begin{equation} \label{TwoPointSolEps2}
\begin{split}
\mathcal{G}_2(\eta) &= \frac{d_1}{\eta} + \frac{d_2}{1 - \eta} +  \frac{ d_3 \log \eta}{1 -  \eta} + \frac{d_4 \log( 1 - \eta)}{\eta} \\
& - \frac{N + 2}{4(N + 8)^2} \left(\frac{ \log \eta}{\eta} \pm \frac{ \log( 1-  \eta)}{1 - \eta} \right) + \frac{ \log^2 \eta}{8 \eta} \pm \frac{ \log^2( 1-  \eta)}{8 (1 - \eta)} \\
& + \frac{(N + 2)^2}{8(N + 8)^2} \left(\frac{ \log^2 (1 - \eta)}{\eta} \pm \frac{ \log^2( \eta)}{1 - \eta} \right) + \frac{N + 2}{ 4 (N + 8)} \left(\frac{1}{\eta} \pm \frac{1}{1 - \eta} \right) \log \eta \log (1 - \eta)  . 
\end{split} 
\end{equation}
To fix the constants, we again use the normalization and demand symmetry/antisymmetry under $\eta \rightarrow 1 - \eta$. This sets $d_1 = d_2 = 0$, and $d_4 = \pm d_3$. To fix $d_3$, we recall that the in the direct channel, $\eta \rightarrow 0$, the correlator should behave as 
\begin{equation}
\begin{split}
\mathcal{G}(\eta) &= \eta^{- \Delta_{\phi} } + \mu_{\phi \phi 0} \eta^{\frac{\Delta_0^{d.t.}}{2} - \Delta_{\phi}} + \textrm{higher orders in $\eta$} \\
&= \eta^{- \Delta_{\phi} } + \mu^{(0)}_{\phi \phi 0}
+ \epsilon \left(\mu^{(1)}_{\phi \phi 0} + \frac{\gamma_0^{(1)}}{2} \log \eta \right) \\
&+ \epsilon^2 \left( \mu^{(2)}_{\phi \phi 0} + (\mu^{(1)}_{\phi \phi 0} \gamma_0^{(1)} + \mu^{(0)}_{\phi \phi 0} \gamma_0^{(2)}) \frac{\log \eta}{2} +   \mu^{(0)}_{\phi \phi 0} (\gamma_0^{(1)})^2 \frac{\log^2 \eta}{8} \right) + O(\eta). 
\end{split}
\end{equation}
Comparing the $\log \eta$ terms at order $\epsilon^2$ with \eqref{TwoPointSolEps2} then tells us
\begin{equation}
\begin{split}
&\mu^{(1)}_{\phi \phi 0} \gamma_0^{(1)} + \mu^{(0)}_{\phi \phi 0} \gamma_0^{(2)} = 2 d_3 - \frac{N + 2}{2 (N + 8)} \\
 \implies & d^+_3 = \frac{3 (N + 2) (3 N + 14)}{2 (N + 8)^3}, \ \ d^-_3 = -\frac{3 (N + 2) (N - 2)}{2(N + 8)^3}. 
\end{split}
\end{equation}
This gives us an explicit expression for the complete two-point function to order $\epsilon^2$ for the Wilson-Fisher model on $\mathbb{RP}^d$. Expanding the two-point function in powers of $\eta$ and extracting the OPE coefficients, it is easy to check that this agrees with the results in \eqref{Epsilon2OPEResults} which  we found by using functionals. The large $N$ limit of this solution also agrees with \eqref{TwoPointCriticalO(N)+-} in $d = 4- \epsilon$.

\section{Comments on relation to bulk reconstruction}\label{Sec:7}
Our discussion of  $\mathbb{RP}^d$ CFTs can also be related to the bulk reconstruction program in the large $N$ limit. In this section we  make a number of comments regarding the connection with previous works.

Let us begin by noticing that inserting a local operator in Euclidean $AdS_{d+1}$ has the same effect of breaking the isometry from $SO(d+1,1)$ to $SO(d,1)$, as performing the $\mathbb{Z}_2$ quotient (\ref{AdScalI}). This is because the $\mathbb{Z}_2$ quotient selects a special fixed point $N_c$, just as inserting a local bulk operator. However, the identification under the inversion does not further change the Lie algebra of the residual symmetry group, and we will not impose such identifications in this section.
Notice that $N_c$ now is no longer a special point, because AdS space is homogenous. Nevertheless, we will always use the AdS isometry generators to move the local bulk operator to $N_c$ without loss of generality, so that it is easier to make a connection with the discussions in Section \ref{Sec:3}. 

 We will compare the holographic objects considered in Section \ref{Sec:3} with those arising from the Hamilton-Kabat-Lifschytz-Lowe (HKLL) approach for constructing local bulk operators \cite{Hamilton:2005ju,Hamilton:2006fh,Hamilton:2006az}, which is perturbative in nature in the $1/N$ expansion.\footnote{There are also intrinsically non-perturbative and state-independent developments which exploit the identification of twisted Ishibashi states with bulk operators \cite{Miyaji:2015fia,Nakayama:2015mva, Verlinde:2015qfa, Nakayama:2016xvw,Goto:2016wme, Lewkowycz:2016ukf}, and are explored most extensively in two dimensions. In fact, twisted Ishibashi states can be considered even when the boundary spacetime does not have crosscap insertions. } In the HKLL approach, a bulk field at a point in AdS can be defined by smearing the CFT operator with the bulk-to-boundary propagator (we do not keep track of the overall normalizations in this section)  
\begin{equation}\label{HKLL}
\Phi^{(0)}_\Delta(N_c)=\int dP\, G_{B\partial}^\Delta(N_c,P) \mathcal{O}_\Delta(P)\;.
\end{equation}
The bulk-boundary two-point function can be obtained by performing the above smearing in the CFT two-point function, and we get 
\begin{equation}
\langle \Phi^{(0)}_\Delta(N_c) \mathcal{O}_\Delta(P)\rangle \propto G_{B\partial}^\Delta(N_c,P)\;.
\end{equation}
This reproduces the one-point function (\ref{AdS1pt}). 

\begin{figure}[htbp]
\begin{center}
\includegraphics[width=0.58\textwidth]{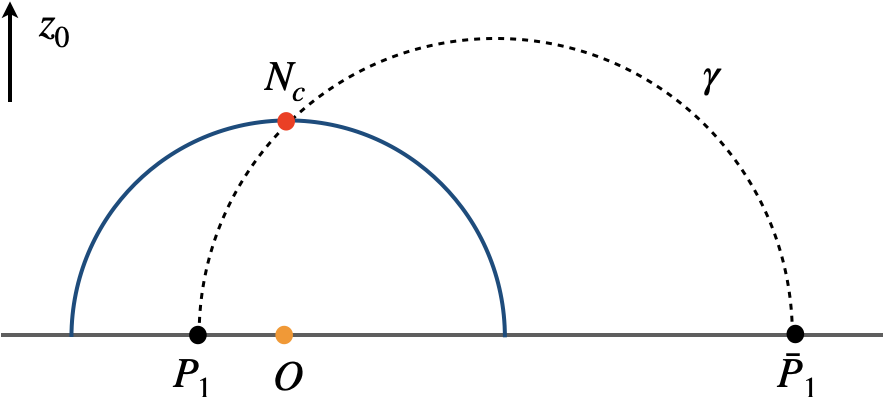}
\caption{Illustration of the branch cut singularity in the conformal block from the geodesic Witten diagram picture. In the limit $\eta\to1$, the point 2 approaches the image of point 1. The geodesic line connecting these two points now goes through $N_c$, and the geodesic Witten diagram integral divergence.}
\label{Fig:Wgeoetaeq1}
\end{center}
\end{figure}

However, applying (\ref{HKLL}) to a CFT three-point function runs into the problem of non-vanishing commutators for space-like separated operators, as the prescription is only good for free particles. Doing the integral, we get \cite{Kabat:2011rz}
\begin{equation}
\langle \Phi^{(0)}_\Delta(N_c) \mathcal{O}_{\Delta_1}(P_1)\mathcal{O}_{\Delta_2}(P_2)\rangle\propto \frac{\eta^{\frac{\Delta-\Delta_1-\Delta_2}{2}} }{(1+x_1^2)^{\Delta_1}(1+x_2^2)^{\Delta_2}}{}_2F_1\left(\tfrac{\Delta+\Delta_1-\Delta_2}{2},\tfrac{\Delta+\Delta_2-\Delta_1}{2};\Delta-\tfrac{d}{2}+1;\eta\right)\;.
\end{equation}
We recognize that this bulk-boundary three-point function is nothing but the conformal block $g_\Delta(\eta)$, which is not surprising from the symmetry point of view.\footnote{One can act on it with the two-particle conformal Casimir operator on the boundary, and use the equation of motion identity for the bulk-to-boundary propagator, to show that it is an eigenfunction.} The conformal block has a branch cut starting at $\eta=1$ where points are space-like separated. The existence of the singularity indicates a failure of the micro-causality. Meanwhile, we recall that in Section \ref{Sec:geoW} we found an alternative geometric representation for the conformal blocks. This gives the above three-point function an interpretation in terms of a geodesic Witten diagram. Using this picture, we can obtain an intuitive understanding of the singularity without computing the integral. We note that the point $\eta=1$ corresponds to the limit where one boundary point is approaching the image of the other boundary point. In this limit, the geodesic line which connects the two boundary points goes through the fixed bulk point $N_c$ (see Figure \ref{Fig:Wgeoetaeq1}). This creates a short distance singularity in the integral, and makes the three-point function singular. 

To resolve the singularity and restore micro-causality, \cite{Kabat:2015swa} proposed that one should correct $ \Phi^{(0)}_\Delta$ with infinitely many double-trace operators
\begin{equation} \label{BulkReconImproved}
\Phi_\Delta(N_c)=\Phi^{(0)}_\Delta(N_c)+\sum_{n=0}^\infty a^{12}_n \int dP\, G_{B\partial}^{\Delta_1+\Delta_2+2n}(N_c,P) :\mathcal{O}_{\Delta_1}\mathcal{O}_{\Delta_2}:(P)
\end{equation}
where $a^{12}_n$ contains a $1/N$ suppression so that both terms contribute at the same order. The coefficients of the double-trace operators can be systematically determined by cancelling the branch point singularity at $\eta=1$ \cite{Kabat:2015swa}. As a result, the three-point function $\langle \Phi_\Delta\mathcal{O}_{\Delta_1}\mathcal{O}_{\Delta_2}\rangle$ now contains not only the single-trace conformal block $g_{\Delta}(\eta)$, but also infinitely many double-trace conformal blocks $g_{\Delta_1+\Delta_2+2n}(\eta)$. One can imagine that all the contributions to $\langle \Phi_\Delta\mathcal{O}_{\Delta_1}\mathcal{O}_{\Delta_2}\rangle$ have been resummed. Then, the end result of this prescription for the bulk reconstruction should coincide with the exchange Witten diagram $W^{exchange}_\Delta$ defined in (\ref{defW}). To see it, we recall that the conformal block decomposition of $W^{exchange}_\Delta$ has the same structure as \eqref{BulkReconImproved}, and $W^{exchange}_\Delta$ is free of singularities at $\eta=1$. Note that there is one detail we have glossed over: there are also homogeneous solutions to $a^{12}_n$ which do not have branch singularities. But these solutions just correspond to contact Witten diagrams, which are polynomials of the cross ratio.   

The above holographic reconstruction of the three-point function $\langle \Phi_\Delta\mathcal{O}_{\Delta_1}\mathcal{O}_{\Delta_2}\rangle$ can be alternatively phrased as a conformal bootstrap problem. We can ask the following question in the spirit of the seminal work \cite{Heemskerk:2009pn}: given the appearance of a single-trace operator with dimension $\Delta$, what is the total contribution to the field theory two-point function of $\mathcal{O}_{\Delta_1}$, $\mathcal{O}_{\Delta_2}$ at order $1/N$, dictated by the partially broken $SO(d,1)\subset SO(d+1,1)$ conformal symmetry?\footnote{This requires the single-trace operator $\mathcal{O}_\Delta$ to appear in the OPE of $\mathcal{O}_{\Delta_1}\times \mathcal{O}_{\Delta_1}$, and also to have a nonzero one-point function. The latter is possible because we assume the conformal symmetry to be partially broken. We  emphasize again that the breaking of conformal symmetry is not due to placing the theory on $\mathbb{RP}^d$ (the space is still $\mathbb{R}^d$), but due to the presence of a bulk local operator (to be interpreted from the solution to the problem).} This question is similar in flavor to the question asked in \cite{Alday:2017gde} about four-point functions in CFTs with full conformal symmetry. Since we are in the large $N$ limit, the conditions of our question indicate that the conformal block decomposition should take the following form
\begin{equation}\label{calGPhiOO}
\begin{split}
\mathcal{G}_{\Phi\mathcal{O}\mathcal{O}}(\eta)={}&\mu g_\Delta(\eta)+\sum_{n=0}^\infty b_n g_{\Delta_1+\Delta_2+2n}(\eta)\\
={}&\sum_{n=0}^\infty c_n \bar{g}_{\Delta_1+\Delta_2+2n}(\eta)
\end{split}
\end{equation}
where only double-trace conformal blocks are allowed to appear in addition to the single-trace conformal block. We can view these expressions as the leading order deformation to the mean field theory two-point function, by adding a single-trace operator.
 In order to get rid of the anticipated ambiguities in the double-trace operators coming from contact diagrams, we should also impose a bound on the Regge behavior\footnote{Similar issues with contact diagrams can also arise in the four-point function problem, and can be eliminated by imposing conditions on the Regge growth.} 
\begin{equation}
|\mathcal{G}_{\Phi\mathcal{O}\mathcal{O}}|\lesssim |\eta|^{-\epsilon}\;,\;\text{when  }\;\; \eta\to\infty\;.
\end{equation}
The extra homogeneous solutions with just double-trace conformal blocks can always be conveniently added back in the very end. This conformal bootstrap problem can then be easily solved by using the analytic functionals which we introduced in Section \ref{Sec:basisfunctional}. Applying the basis of functionals on (\ref{calGPhiOO}), we find that 
\begin{equation}
b_n=-\mu\,\omega_n(g_\Delta)\;,\quad c_n=\mu\,\omega_n(\bar{g}_\Delta)\;.
\end{equation}
Comparing with (\ref{Ping}), this indicates that $\mathcal{G}_{\Phi\mathcal{O}\mathcal{O}}(\eta)$ is just proportional to the uniquely defined Polyakov-Regge block $\mathcal{P}_\Delta(\eta)$, {\it i.e.}, an exchange Witten diagram with a local operator in the bulk AdS.

\section{Future directions}\label{Sec:8}
In this paper we performed an analytic study of CFTs on real projective space. We gave a detailed account of a toy model of holography on a $\mathbb{Z}_2$ quotient of AdS, and studied  properties of Witten diagrams on this background. The investigation led to a basis of analytic functionals dual to double-trace conformal blocks. We explicitly constructed these functionals from the conformal block decomposition coefficients of exchange Witten diagrams. Although the functionals stem from a toy holography model, they apply universally to $\mathbb{RP}^d$ CFTs. In particular, we applied these functionals to study $O(N)$ vector model in $4-\epsilon$ expansion, and obtained one-point functions to order $\epsilon^2$. We also studied in detail the large $N$ $O(N)$ vector model on $\mathbb{RP}^d$ using independent field theory techniques, and obtained results that are consistent with the $\epsilon$-expansion. Our work leads to a number of interesting future directions.

An interesting extension of our work is to include fermions, and study models  on real projective space such as QCD and the Gross-Neveu model. Including fermions is also necessary for considering theories with supersymmetry. The case of 4d $\mathcal{N}=4$ SYM on $\mathbb{RP}^4$ has been recently considered in \cite{Wang:2020jgh} using supersymmetric localization techniques. It will be nice to study it using other analytic techniques, such as those developed in this paper. 

Another related direction to explore is thermal CFTs obtained by compactification on $S^1 \times \mathbb{R}^{d- 1}$ \cite{Iliesiu:2018fao}, where two-point functions also nontrivially depend on the spacetime coordinates. Similar to our setup, the new data of thermal CFTs enters as the one-point function coefficients. But unlike our case, spinning operators are also allowed to have non-vanishing thermal one-point functions. Furthermore, the conformal symmetry is fully broken to $U(1) \times O(d-1)$ in thermal CFTs, and two-point functions depend on two independent cross ratios instead of one. Despite the differences, it would be interesting to see if some of our techniques can be generalized to study that problem.

As we pointed out in Section \ref{Sec:5}, the existence of the two-term dimensional reduction formula for conformal blocks suggests an extension of the Parisi-Sourlas supersymmetry to real projective space. It would be very interesting to study in detail the realization of the symmetry in concrete models such as branched polymers, and test its  equivalence with the Yang-Lee critical theory on a real projective space with two dimensions less.

A noticeable omission in the literature of $\mathbb{RP}^d$ CFTs is the top-down construction of their holographic duals. On the other hand, theories such as $\mathcal{N}=4$ SYM are completely well-defined on $\mathbb{RP}^4$ at weak coupling, and presumably will remain well-defined at strong coupling as well. Finding a dual description in IIB supergravity for the strong coupling limit  should therefore be possible. It will be interesting to find such explicit backgrounds, which will provide the starting point for doing holographic calculations. Similarly, it would be interesting to investigate the same question for the Vasiliev higher-spin theory and further check the conjectured duality to $O(N)$ vector model \cite{Klebanov:2002ja} by using the results obtained in this paper.

Related to studying the holographic duals, an interesting question to ask is whether there are any universal results that can be derived for double-trace deformation of CFT on $\mathbb{RP}^d$ similar in spirit to \cite{Witten:2001ua, Gubser:2002zh, Gubser:2002vv, Hartman:2006dy}. We study the two-point function and free energy in the large $N$ critical $O(N)$ vector model, which can be obtained as a double-trace deformation of the free $O(N)$ model. It would be interesting to see if some of the results are model independent and hold true for more general double-trace deformation.

\section*{Acknowledgments}
The work of S.G and H.K is supported in part by the US NSF under Grants No. PHY-1620542 and PHY-1914860. The work of X.Z. is supported in part by the Simons Foundation Grant No. 488653.

\appendix

\section{$\mathbb{RP}^d$ free energy} \label{AppendixFreeEnergy}
In this appendix, we show how to compute the $\mathbb{RP}^d$ free energy for the critical $O(N)$ vector model. To calculate the free energy, we need to go to a compact space, so we will do this on the $\mathbb{Z}_2$ quotient of sphere. The action on the sphere for the $O(N)$ model is \footnote{We are using a different normalization of $\phi$ in this appendix from the rest of the paper such that the two-point function goes like \eqref{NormalizationAppendix}. } 
\begin{equation}
S = \frac{1}{2} \int d^d x \sqrt{g} \left( (\partial_{\mu} \phi^I)^2 + \frac{d (d - 2)}{4} \phi^I \phi^I + \sigma \phi^I \phi^I   \right).
\end{equation}
As we saw earlier \eqref{1ptfunSphere}, the one-point function of $\sigma$ is just going to be a constant. At leading order in large $N$ the effect of $\sigma$ will only be through this one-point function, so it is just like $N$ free massive fields on the sphere and the action becomes quadratic. To calculate the free energy, we need to calculate the determinant of this quadratic operator. For that purpose, we need to study the behavior of eigenfunctions of the scalar Laplacian and calculate the degeneracies of the eigenfunctions which are odd or even under the $\mathbb{Z}_2$ quotient. The eigenfunctions of the scalar Laplacian on a $d$ dimensional sphere are spherical harmonics $Y_{\vec{l}} (\vec{\theta})$ with $\vec{l} = \{l_1, \ldots l_d \}$ satisfying $|l_1| \leq l_2 \leq \ldots l_d$. The eigenvalues and degeneracy is given by 
\begin{equation}
\Box_d Y_{\vec{l}} (\vec{\theta}) = - l_d (l_d + d - 1), \ \ \ \mathrm{dim}(l_d) = \frac{\Gamma\left(d + l_d + 1 \right)}{\Gamma\left(d + 1 \right) \Gamma\left(l_d + 1 \right)} - \frac{\Gamma\left(d + l_d - 1 \right)}{\Gamma\left(d + 1 \right) \Gamma\left(l_d - 1 \right)}.
\end{equation}
The eigenfunctions $Y_{\vec{l}} (\vec{\theta})$ have explicit construction in terms of associated Legendre polynomials \cite{doi:10.1063/1.527513}. Under parity, they behave as $Y_{\vec{l}} (-\vec{\theta}) = (-1)^{l_d} Y_{\vec{l}} (\vec{\theta})$ \cite{Frye:2012jj}. So to compute the free energy, we just need to sum over either even or odd values of $l_d$ depending upon whether we choose to identify the scalar with itself or minus itself. So for instance, for $+$ parity, we need to perform the sum 
\begin{equation} \label{FreeEnergyPlus}
F^{+}(a_{\sigma}) = \frac{N}{2} \sum_{l_d \in 2 \mathbb{Z}} \mathrm{dim}(l_d) \log \left( l_d (l_d + d - 1) + \frac{d (d - 2)}{4} + \frac{ a_{\sigma}}{4} \right)
\end{equation}
while for $-$ parity, we have the exact same sum but over $l_d \in 2 \mathbb{Z} + 1$.

Let's first consider the case of free theory, when $a_{\sigma} = 0$. Then there is a more convenient way of writing this sum as 
\begin{equation}
F^{+}(\Delta) = \frac{N}{2} \sum_{l_d \in 2 \mathbb{Z}} \mathrm{dim}(l_d) \log \frac{\Gamma\left( l_d + d - \Delta \right)}{\Gamma \left( l_d + \Delta \right)}
\end{equation}
which goes back to the previous expression for $\Delta = d/2 - 1$. Note that this sum vanishes for $\Delta = d/2$. To do this sum, we can take a derivative, perform the sum and then integrate it back \cite{Diaz:2007an, Giombi:2014xxa}
\begin{equation}
\frac{\partial F^{+}}{\partial \Delta} = \frac{N}{4} \left[ \Gamma(-d) (d - 2 \Delta) \left( \frac{\Gamma(d - \Delta)}{\Gamma(1 - \Delta)} - \frac{\Gamma(\Delta)}{\Gamma(1 + \Delta - d)} \right) + \frac{ \Gamma(\Delta) \Gamma(d - \Delta)}{\Gamma(d)} \right]
\end{equation} 
where we had to use the integral representation of polygamma function to perform the sum. Similar method also works for $-$ parity. So for any $\Delta$, we get 
\begin{equation}
\begin{split}
F^{\pm} (\Delta) = \frac{N}{4}  \int_{0}^{\Delta - \frac{d}{2}} d u \bigg[& - 2 \ \Gamma(-d) u \left( \frac{\Gamma \left( \frac{d}{2} - u \right)}{\Gamma\left( 1 - u -  \frac{d}{2}\right)} - \frac{\Gamma \left( \frac{d}{2} + u \right)}{\Gamma\left( 1 + u -  \frac{d}{2}\right)} \right) \\
&\pm  \frac{ \Gamma \left( \frac{d}{2} + u \right) \Gamma \left( \frac{d}{2} - u \right)}{\Gamma(d)} \bigg].
\end{split}
\end{equation} 
Here we are only interested in the case of $\Delta = d/2 - 1$ which gives 
\begin{equation}
F^{\pm}  = - \frac{N}{4}  \int_{0}^{1} d u  \frac{ \Gamma \left( \frac{d}{2} + u \right) \Gamma \left( \frac{d}{2} - u \right)}{\Gamma(d)}  \left( \frac{2 u \sin \pi u}{d \sin \left( \frac{\pi d}{2} \right)}  \pm 1  \right).
\end{equation}
As an explicit example, in $d=3$ we find
\begin{equation}
F^{\pm}  = N \int_0^1 du 
\frac{\pi}{96}   \left(1-4 u^2\right) (2 u \tan (\pi  u) \mp 3 \sec (\pi  u))
=N\left(\frac{\log(2)}{16}-\frac{3\zeta(3)}{32\pi^2}\mp \frac{K}{2\pi}\right)
\end{equation}
where $K$ is Catalan's constant. 

As a consistency check, note that adding the results for both parities gives us back the well-known sphere free energy of a free scalar on a sphere \cite{Klebanov:2011gs}. 

For the interacting case, the sum is hard to perform, but it can be performed if we take a derivate with respect to $a_{\sigma}$ of eq. \eqref{FreeEnergyPlus} first
\begin{equation}
\frac{\partial F^{+}(a_{\sigma})}{\partial a_{\sigma}} = \frac{N}{8} \sum_{l_d \in 2 \mathbb{Z}} \frac{\mathrm{dim}(l_d)}{  l_d (l_d + d - 1) + \frac{d (d - 2)}{4} + \frac{ a_{\sigma}}{4} }. 
\end{equation}
This sum can then be performed and after some manipulations involving Hypergeometric identities, we get 
\begin{equation} \label{FreeEnergySum+}
\begin{split}
&\frac{\partial F^{+}(a_{\sigma})}{\partial a_{\sigma}} = \frac{ N 2^{-1-d}}{d(d-2) + a_{\sigma}}-\frac{N}{64} \sqrt{\pi } d \Gamma (1-d)\bigg[\Gamma \left(\frac{d - 1 - \sqrt{1-a_{\sigma}}}{4} \right)  \\
& \times \, _3\tilde{F}_2\left(\frac{1-d}{2},1-\frac{d}{2},\frac{d + 3 - \sqrt{1-a_{\sigma}}}{4} ;\frac{3}{2},\frac{-3 d + 7 - \sqrt{1-a_{\sigma}}}{4} ;1\right) +\Gamma \left(\frac{d - 1 + \sqrt{1-a_{\sigma}}}{4} \right)  \\
&\times \, _3\tilde{F}_2\left(\frac{1-d}{2},1-\frac{d}{2},\frac{d + 3 + \sqrt{1-a_{\sigma}}}{4} ;\frac{3}{2},\frac{-3 d + 7 + \sqrt{1-a_{\sigma}}}{4} ;1\right)\bigg].
\end{split}
\end{equation}
Now we should impose $\frac{\partial F^{+}(a_{\sigma})}{\partial a_{\sigma}} = 0$ for the critical theory, because at large $N$ we expect the $\sigma$ path integral to be dominated by the saddle point of free energy. It can be checked that this happens for $a_{\sigma} = - (d - 2)(d - 4)$, which precisely agrees with what we found in \eqref{TwoPointCriticalO(N)+-}. For the $-$ parity, it is the same sum, but over odd integers, and it gives 
\begin{equation} \label{FreeEnergySum-}
\begin{split}
&\frac{\partial F^{-}(a_{\sigma})}{\partial a_{\sigma}} = -\frac{N 2^{-1-d} (d + 1)}{(d-1) (d (d+2)+x)} - \frac{N \sqrt{\pi } \ d (d + 1) \ \Gamma (1-d)}{128} \bigg[ \Gamma \left(\frac{d + 1 - \sqrt{1-a_{\sigma}}}{4}\right) \\
& \times \, _3\tilde{F}_2\left(1-\frac{d}{2},\frac{3-d}{2},\frac{d + 5 - \sqrt{1-a_{\sigma}}}{4};\frac{5}{2},\frac{ -3 d + 9 - \sqrt{1-a_{\sigma}}}{4}; 1 \right) +\Gamma \left(\frac{d + 1 + \sqrt{1-a_{\sigma}}}{4}\right) \\
&\times \, _3\tilde{F}_2\left(1-\frac{d}{2},\frac{3-d}{2}, \frac{d + 5 + \sqrt{1-a_{\sigma}}}{4};\frac{5}{2},\frac{ -3 d + 9 + \sqrt{1-a_{\sigma}}}{4}; 1 \right) \bigg].
\end{split}
\end{equation}
It can again be checked that this vanishes for $a_{\sigma} = - (d - 4)(d - 6)$. There is another way to arrive at this result for the free energy. The derivative of the free energy with $a_{\sigma}$ is related to the one point function of $\phi^I \phi^I$ as 
\begin{equation}
\frac{\partial F^{\pm}(a_{\sigma})}{\partial a_{\sigma}} = \frac{\textrm{Vol}(S^d)}{2} \frac{\langle \phi^I \phi^I \rangle}{8} ,\hspace{1cm} \textrm{Vol}(S^d) = \frac{2 \pi^{\frac{d + 1}{2}}}{\Gamma\left( \frac{d + 1}{2} \right)}
\end{equation}
where we used the fact that the volume of $S^d/\mathbb{Z}_2$ is half the volume of $S^d$. But this one-point function can also be obtained from the coincident point limit of the two-point function of $\phi$. Since at large $N$, $\phi$ is equivalent to a massive free field on a sphere, the two-point function is just the Green's function on a sphere for a massive scalar \cite{allen1986}\footnote{The result in \cite{allen1986} is given in terms of the the geodesic distance $\mu$, which is related to $\eta$ as $\eta = \sin^2 (\mu/2)$. }. We already encountered this large $N$ two-point function in \eqref{LargeNTwoPointSol} which gives
\begin{equation}
\begin{split}
\mathcal{G}(\eta) =& \frac{C_1}{(1 - \eta)^{\frac{d}{2} - 1}} \ {}_2F_1 \left( \frac{1 + \sqrt{1 - a_{\sigma}}}{2}, \frac{1 - \sqrt{1 - a_{\sigma}}}{2},  2 - \frac{d}{2}, 1 - \eta \right) \\
&+  \frac{C_2}{\eta^{\frac{d}{2} - 1}} \ {}_2F_1 \left( \frac{1 + \sqrt{1 - a_{\sigma}}}{2}, \frac{1 - \sqrt{1 - a_{\sigma}}}{2},  \frac{d}{2}, 1 - \eta \right).
\end{split}
\end{equation}
When we perform the $Z_2$ quotient of the sphere, we have to impose the crossing symmetry requirement $\mathcal{G}(\eta) = \pm \mathcal{G}(1 - \eta)$ which implies the relation \eqref{LargeNTwoPointCoeff} for the coefficients. Here we are using a different normalization, so the overall constant can be fixed by requiring that 
\begin{equation} \label{NormalizationAppendix}
\mathcal{G}(\eta) = \frac{\Gamma \left( \frac{d}{2} - 1 \right)}{(4 \pi)^{d/2}} \frac{1}{\eta^{\frac{d}{2} - 1}}, \hspace{1cm} \eta \rightarrow 0.
\end{equation}
The one-point function of $\phi^I \phi^I$ can then be read off from the constant piece in the small $\eta$ expansion of $\mathcal{G}(\eta)$. That gives
\begin{equation} \label{FreeEnergyIntE}
\frac{\partial F^{\pm}(a_{\sigma})}{\partial a_{\sigma}} =  \frac{ \pm \pi ^{\frac{d+3}{2}} \Gamma \left(\frac{d}{2}-1\right) \Gamma \left(2-\frac{d}{2}\right)}{8 (4\pi)^{\frac{d}{2}} \Gamma \left(\frac{d+1}{2}\right) \Gamma \left(\frac{d}{2}\right) \Gamma \left(\frac{3 - d - \sqrt{1 - a_{\sigma}}}{2}\right) \Gamma \left(\frac{3 - d + \sqrt{1 - a_{\sigma}}}{2} \right) \left(\sin \left(\frac{\pi  d}{2}\right) \mp \cos \left(\frac{\pi  \sqrt{1 - a_{\sigma}}}{2}\right)\right)}. 
\end{equation}
Note that the large $N$ saddle point requirement of vanishing of the derivative of the free energy is then clearly equivalent to the requirement that the operator $\phi^2$ of dimension $d - 2$ is replaced by operator $\sigma$ of dimension $2$, which is what we used in subsection \ref{Sec:testfunctionals}. We were not able to show analytically that these formulas for the derivative of free energy are the same as \eqref{FreeEnergySum+} and \eqref{FreeEnergySum-}, but we checked numerically over a range of values of $d$ and $a_{\sigma}$ that they agree. We can also see that they also have the same zeroes as a function of $a_{\sigma}$. 

We can then find the value of the free energy at the critical point by integrating these expressions. In $d=3$, we get
\begin{equation}
\begin{split}
F^{+}_{\textrm{crit}} &= F^{+}(a_{\sigma} = 0) + \int_{0}^{- (d - 2)(d - 4)} d a_{\sigma} \ \frac{\partial F^{+}(a_{\sigma})}{\partial a_{\sigma}} \ \xrightarrow{\text{$d = 3$}} \\
&= F^{+}(a_{\sigma} = 0) + \frac{N (16 \pi K- 21 \zeta (3) - 2 \pi ^2 \log (2)) }{32 \pi ^2}   = - \frac{3 N }{4 \pi^2} \zeta(3) \\
F^{-}_{\textrm{crit}} &= F^{-}(a_{\sigma} = 0) + \int_{0}^{- (d - 4)(d - 6)} d a_{\sigma} \ \frac{\partial F^{-}(a_{\sigma})}{\partial a_{\sigma}} \ \xrightarrow{\text{$d = 3$}}   \\
& =  F^{-}(a_{\sigma} = 0) + \frac{N (-16 \pi K -21 \zeta (3) + 14 \pi^2 \log (2))}{32 \pi ^2} = - \frac{3 N }{4 \pi^2} \zeta(3) + \frac{N \log(2)}{2}  .
\end{split}
\end{equation}
Note that both in the free and interacting theory, we get $F^{-} > F^{+}$ in $d = 3$. We can also do a similar computation in $d = 4 - \epsilon$
\begin{equation}
\begin{split}
F^{+}_{\textrm{crit}} &= F^{+}(a_{\sigma} = 0) + \int_{0}^{2 \epsilon} d a_{\sigma} \ \frac{\partial F^{+}(a_{\sigma})}{\partial a_{\sigma}} \ \xrightarrow{\text{$d = 4 - \epsilon$}}  \ = F^{+}_{\textrm{free}} +  \frac{N \epsilon}{96} + O(\epsilon^2) \\
F^{-}_{\textrm{crit}} &=  F^{-}(a_{\sigma} = 0) + \int_{0}^{ - 2 \epsilon} d a_{\sigma} \ \frac{\partial F^{-}(a_{\sigma})}{\partial a_{\sigma}} \ \xrightarrow{\text{$d = 4 - \epsilon$}} \ = F^{-}_{\textrm{free}} + \frac{N \epsilon}{96} + O(\epsilon^2).
\end{split}
\end{equation}
We can check this against a computation in $\epsilon$ expansion. Using the action in \eqref{ActionSpherePhi4}, we get
\begin{equation}
F^{\pm}_{\textrm{crit}} = F^{\pm}_{\textrm{free}} + \frac{\lambda}{4} \int_{S^d/\mathbb{Z}_2} d^d x \sqrt{g} \langle (\phi^I \phi^I)(\phi^K \phi^K)(x)\rangle_{} \rangle =  F^{\pm}_{\textrm{free}} + \frac{N (N + 2) \epsilon}{96 (N + 8)}
\end{equation}
which agrees at large $N$ with the large $N$ expansion result. 
\bibliography{RPdCFT} 
\bibliographystyle{utphys}

\end{document}